\documentclass[manuscript]{acmart}

\usepackage{amsmath}
\usepackage{mathtools}
\usepackage{graphicx}
\usepackage{adjustbox}
\usepackage{pdflscape}
\usepackage[T1]{fontenc}  
\usepackage{subcaption}
\usepackage{multirow}
\usepackage{xcolor}
\usepackage{float}
\usepackage{tikz}
\usetikzlibrary{mindmap, shadows}
\usepackage{array}
\usepackage{tabularx}
\usepackage{longtable}
\usepackage{booktabs}
\usepackage{xltabular}
\usepackage{tcolorbox}

\AtBeginDocument{%
  }

\setcopyright{none}
\copyrightyear{2027}
\acmYear{2027}
\acmDOI{}
\acmConference{}{}{}
\acmISBN{}
\graphicspath{{Figures/}}

\begin{document}
\raggedbottom
\title[LLM-Based Family Education: A Scoping Review]{Characterizing LLM-Based Family Education through the Lens of Activity Theory: A Scoping Review of the HCI Literature}

\author{Lan Luo}
\affiliation{%
  \institution{The Hong Kong University of Science and Technology (Guangzhou)}
  \city{Guangzhou}
  \country{China}}
\email{lluo476@connect.hkust-gz.edu.cn}

\author{Yuqi Liang}
\affiliation{%
  \institution{The Hong Kong University of Science and Technology (Guangzhou)}
  \city{Guangzhou}
  \country{China}}
\email{lluo476@connect.hkust-gz.edu.cn}

\author{Jie Cai}
\affiliation{%
  \institution{Nan Kai university}
  \city{Tianjin}
  \country{China}}
\email{lluo476@connect.hkust-gz.edu.cn}

\author{Anqi Wang}
\affiliation{%
  \institution{The Hong Kong University of Science and Technology}
  \city{Hongkong,SAR}
  \country{China}}
\email{awangan@connect.ust.hk}

\author{Dongyijie Primo Pan}
\affiliation{%
  \institution{The Hong Kong University of Science and Technology (Guangzhou)}
  \city{Guangzhou}
  \country{China}}
\email{dpan750@connect.hkust-gz.edu.cn}

\author{Muzhi Zhou}
\affiliation{%
  \institution{The Hong Kong University of Science and Technology (Guangzhou)}
  \city{Guangzhou}
  \country{China}}
\email{mzzhou@hkust-gz.edu.cn}

\author{Chun YU}
\affiliation{%
  \institution{Tsinghua University)}
  \city{Beijing}
  \country{China}}

\author{Pan Hui}
\affiliation{%
  \institution{Hong Kong University of Science and Technology}
  \city{Hong Kong}
  \country{China}}
\affiliation{%
  \institution{The Hong Kong University of Science and Technology (Guangzhou)}
  \city{Guangzhou}
  \country{China}}
\email{panhui@ust.hk}

\begin{abstract}
Large language models (LLMs) are increasingly involved in family education, yet HCI has not systematically explained the educational interactions that emerge around them. This scoping review analyzes 53 HCI studies from 6,540 records across 19 venues. Using activity theory and AODM, it relates participants and educational objects to mediation, labour, and rules. We find that the literature centers on child--parent interaction and on language, AI literacy, and relational learning. The introduction of LLMs enabled conversational, embodied, and spatial systems to generate support from the context of an unfolding interaction. LLMs redistributed educational labour, while family and institutional rules left parents and professionals responsible for interpreting outputs and deciding how they entered practice. Evidence across families and educational purposes remains limited, especially on sustained personalization, repair labour, and how families negotiate authority and rules. The review offers a framework explaining how LLM capabilities become organized through family participation.
\end{abstract}

\ccsdesc[500]{Applied computing~Computer-assisted instruction}
\ccsdesc[500]{Human-centered computing~Interactive systems and tools}
\ccsdesc[300]{Human-centered computing~Empirical studies in HCI}

\keywords{Human-centered Computing, Interactive Systems and Tools, Large Language Models, Parental Involvement}

\maketitle

\section{Introduction}
Family education includes children’s learning with family members~\cite{marjoribanks2005family} and parental learning within family life~\cite{euteneuer2014family}. For children’s learning, parents provide guidance through their direct involvement in children’s education~\cite{Sivabalan2024}. Other family members make educational knowledge and support available to children through family interaction~\cite{marjoribanks2005family}. For example, siblings teach and guide one another during everyday family activities~\cite{davies2019sticky,maynard2002cultural}, and children and grandparents learn together through intergenerational co-creation~\cite{kim2025bridging}. For parental learning, raising children provides experiences through which parents develop and revise their understanding of parenting~\cite{euteneuer2014family}. Family education develops through interactions within and across generations. Technology can become part of these interactions, creating new ways for family members to learn together.

Activity theory provides a lens for explaining how LLMs become involved in family education. It takes the activity system as the unit of analysis and examines how people pursue an object through a mediating tool within a socially organized setting~\cite{engestrom2014learning,kaptelinin2012activity}. This perspective fits family education because learning is carried forward through the actions of family members whose participation responds to one another~\cite{hoover1997parents,takeuchi2011new}. When an LLM enters such an activity, it mediates the relation between what participants do and the educational object toward which their activity is directed~\cite{engestrom2014learning}. The role of the system must be understood in relation to how the activity develops and who carries it forward~\cite{nardi1996activity,kaptelinin2009acting}. Activity theory makes this relationship available for systematic analysis.

HCI research has begun to show that LLMs can participate at different levels of family educational interaction. At the level of a shared activity, an LLM can generate stories or questions that parents and children discuss together~\cite{chen2025characterizing,he2025storypal}. Its involvement extends when a system interprets a child’s activity and turns that interpretation into guidance for a parent’s response~\cite{choi2025aacesstalk,dangol2025want}. In homeschooling, this mediation becomes part of longer-term decisions about how learning should reflect curricular expectations and family values~\cite{rifat2026homeroom}. This progression shows how LLM participation can extend from supporting an immediate learning task to mediating how families interpret and organize education together. However, existing HCI research has not yet offered a systematic account of LLM-based family education as a broader research area. Each study explains an interaction in relation to a particular system and educational context, leaving unclear how the different forms of interaction relate to one another. Consequently, the field lacks a shared framework for explaining what patterns of educational interaction emerge when LLM-based technologies enter family education.

To address this gap, this review operationalizes activity theory through the Activity-Oriented Design Method (AODM)~\cite{mwanza2002towards}. AODM provides a coding structure for identifying the object each family educational activity targets and examining how an LLM mediates its pursuit. It then situates this mediation within the activity's social organization, allowing the review to trace how an LLM's role relates to how participation and responsibility are arranged. This review synthesizes 53 HCI studies on LLM-based family education across four areas. The first area concerns the educational objects and outcomes these studies address. The second examines the family subjects, external communities, and divisions of labor involved in LLM-based family education. The third focuses on how LLM-based tools mediate relationships among family members, educational objects, and external expertise. The fourth examines household, institutional, platform, and cultural rules that shape responsible LLM use in family education.

This review makes three contributions to HCI research on LLM-based family education:
\begin{enumerate}
\item \textbf{Systematic synthesis.} We classify and synthesize existing HCI research through a framework informed by activity theory. The classification begins by identifying the family educational activity represented in each study, including who participates and what educational object is pursued. It then examines how an LLM-based system's capabilities and implementation mediate that object. The final stage analyzes how the activity is organized through divisions of labor and rules. This structure provides a systematic account of what has been studied and where the available evidence remains limited.

\item \textbf{Research gaps.} Research gaps are derived from limitations and tensions identified within and between the components of family educational activity. These gaps expose where current evidence does not yet explain how system support relates to the evaluation and organization of family educational activity.

\item \textbf{Implications for HCI.} A grounded research agenda is derived from the limitations identified across the reviewed studies. The resulting agenda addresses broader representations of family education, closer alignment between technical capabilities and educational activities, support for changing forms of labor, and rules that can be negotiated across household and external contexts.
\end{enumerate}

\section{Related Work}
\label{background}

This section develops the conceptual basis for examining the patterns of educational interaction that emerge when LLMs become involved in family education. Section~\ref{sec:family-education-llm} establishes the phenomenon of LLM-supported family education. Within this section, Section~\ref{sec:family-education} defines family education through interdependent family participation, distinguishing it from independent learning at home, while Section~\ref{sec:mediation-forms} reviews the history of technologies used to support family education before the emergence of LLMs and considers how LLM capabilities could change their participation in family educational interaction. It organizes this history around three overlapping strands: conversational systems, physically embodied systems, and spatial or immersive systems. Section~\ref{sec:prior-reviews} then examines prior reviews of technology in family and educational contexts and establishes the need for a synthesis of how family members interact with LLM-based systems to pursue educational goals. Finally, Section~\ref{sec:activity-theory} introduces activity theory as the analytical lens for this synthesis and describes its operationalization through AODM.
\subsection{Family Education Through Cooperative Interaction Using LLMs}
\label{sec:family-education-llm}
\subsubsection{Family Education Through Cooperative Interaction}
\label{sec:family-education}

Family education happens in everyday family life, within family relationships, and through recurring interactions that support development~\cite {mollenhauer1975familienerziehung}. 
It overlaps with informal learning through everyday experience and self-directed exploration~\cite{marsick2015informal,national2009learning} and is distinguished by family members' meaningful involvement in organizing or participating in learning. Parents guide children as they learn and adjust their support in response to children's needs~\cite{hoover1997parents,hill2009parental}. Siblings teach one another through everyday interaction~\cite{davies2019sticky,maynard2002cultural}, while children and caregivers may learn together through shared engagement with media and technology~\cite{takeuchi2011new}. 

The meaningful involvement of family members makes family education a cooperative practice~\cite{hoover1997parents,takeuchi2011new}. Guidance from one family member creates conditions for another member's learning, while the learner's response shapes how the interaction proceeds~\cite{takeuchi2011new}. Educational participation is interdependent within the family. This interdependence matters for HCI because technology enters relationships that already support learning~\cite {takeuchi2011new}. Its role depends on the support it provides and on how family members respond to its use~\cite{beneteau2019communication,sciuto2018hey}. In this review, we treat a study related to family education as one in which a family relationship meaningfully shapes how an educational interaction involving an LLM-based system is organized. This definition includes activities in which family members organize or participate in another member's learning, learn together, or acquire knowledge and skills intended to support another family member. It encompasses relationships involving children aged 0--17 as well as educational interactions among adult family members, including parents and adult children or in-law family relationships.

\subsubsection{Earlier Technological Support before LLMs in Family Education}
\label{sec:mediation-forms}
LLMs enable educational support to be generated as an interaction unfolds~\cite{brown2020language,openai2022chatgpt,yan2024practical}. Instruction prompting allows users to state the educational task in natural language~\cite{brown2020language}. Few-shot prompting then provides examples of how to carry out that task without additional training~\cite{brown2020language}. As the dialogue proceeds, conversation history connects each new contribution to the preceding exchange~\cite{openai2022chatgpt}. Multimodal input brings visual material into the same contextual process~\cite{openai2023gpt4}. These capabilities concern how support is produced. To organize the history preceding LLMs, we discuss three overlapping strands of prior HCI research: \textbf{conversational systems}, \textbf{physically embodied systems}, and \textbf{spatial or immersive systems}. These strands describe different ways technology has participated in educational interaction and are not mutually exclusive; a single system may combine more than one form.

\textbf{Conversational systems} supported spoken or text based interaction~\cite{beneteau2019communication,sciuto2018hey,xu2020exploring}. Before LLMs, these systems commonly operated within a dialogue space specified in advance by their designers~\cite{xu2020exploring,du2021alexa,xu2023rosita}. They matched a user's speech to predefined intents and returned a prewritten response according to a task-specific dialogue sequence. StoryBuddy marked a transition toward generative systems~\cite{zhang2022storybuddy}. It used a BART-based model to generate reading questions from storybooks. During interaction, however, Google Dialogflow~\cite{googleDialogflow} matched children's answers to predefined intents and selected responses according to configured rules. When these systems failed to recognize speech or match a response to a predefined intent, parents and children repeated or rephrased what they said so that the interaction could continue~\cite{beneteau2019communication,du2021alexa,sciuto2018hey}. Adaptation remained bounded by the designed dialogue flow.

\textbf{Physically embodied systems} placed robots within the physical setting in which learning occurred. Bodily cues allowed robots to direct attention and coordinate turns~\cite{scassellati2018teaching}. However, their language and educational roles were commonly specified in advance. Families formed expectations about which roles an in-home robot should assume~\cite{cagiltay2020investigating}. Parents also interpreted or adjusted a robot's guidance when it did not fit the child or the learning situation~\cite{ho2024s}.

\textbf{Spatial or immersive systems} made educational content part of the physical or simulated environments family members interacted with. Earlier AR systems made otherwise hidden processes available for parents and children to examine through shared interaction. In an augmented circuit exhibit, parents and children manipulated physical circuits while viewing a simulation of electron flow~\cite{beheshti2017wires}. ARMath brought spatial mediation into everyday settings by recognizing familiar objects and visualizing their mathematical attributes. Children used these objects as manipulatives for contextual arithmetic and geometry problems, although the evaluation did not examine family participation~\cite{kang2020armath}.

These overlapping strands show how earlier technologies participated in family education through language-based interaction, physical embodiment, and spatial mediation. However, how interactions between family members and LLM-based systems support educational goals remains insufficiently understood. This review examines how family members engage with LLM-based systems in educational interaction.

\subsection{Prior Reviews of Technology in Family and Educational Contexts}
\label{sec:prior-reviews}

Prior reviews of technology use in family contexts have examined how participation is organized around shared digital resources. Ewin et al. analyze the support parents and children exchange during joint mobile media use~\cite{ewin2021impact}. Yu et al. develop an HCI framework for joint engagement around media~\cite{yu2024joint}, while Esteban-Guitart et al. review methods for studying learning in digital home environments~\cite{esteban2026family}. Garg et al. reviewed voice-based conversational agents that supported children's learning and were used in home and family contexts~\cite{garg2022last}. However, these reviews leave unclear how effectively family members use LLM-based systems for educational purposes.

LLMs have assumed increasingly varied roles in supporting children's learning. Yan et al. identify the educational tasks automated by language models and examine the practical and ethical consequences of that automation~\cite{yan2024practical}. Reviews focused on K--12 education show how these capabilities enter particular learning settings. Alfarwan examines how generative AI has been implemented and evaluated in teaching and learning~\cite{alfarwan2025generative}. Among the 30 studies in that review~\cite{alfarwan2025generative}, only Han et al. included parents~\cite{han2024teachers}, leaving family participation peripheral to its account. Lin and Tan draw on a broader corpus spanning formal and informal K--12 settings. Their framework relates learning goals to activity patterns, human--AI roles, and outcomes~\cite{lin2025systematic}. Parents appear within some configurations, but learners aged 3--18 remain the central unit of analysis. A child-centered review brings collaborative participation into view. Cai et al. synthesize how AI systems contribute to children's co-creative processes and propose collaboration with peers and family members as a future design direction~\cite{cai2025child}. Family relationships are not examined as the unit organizing co-creation.

Within educational technology research, Zhang et al. review empirical studies of generative AI in early childhood education from the perspectives of children, families, and teachers~\cite{zhang2026applications}. Their findings show that benefits for young children depend on adult mediation. The review also reports changes in parental effort and involvement, demonstrating that family participation conditions how generative AI contributes to learning. However, HCI research still lacks a cross-study account of how family members interact with LLM-based systems to pursue educational goals in family contexts. Examining this interaction across studies requires an analytical lens that relates the use of LLM-based systems to family participation and educational goals. The following section introduces activity theory as this analytical lens.

\subsection{Activity Theory as a Lens for LLM-based Family Education}
\label{sec:activity-theory}

Activity theory provides an analytical lens for this examination. Rather than defining a technology only by its interface or capabilities, activity theory examines it as a mediating tool whose contribution is directed toward an educational object and organized through participation~\cite{engestrom1987learning,kaptelinin1997activity}. This account of mediation originates in Vygotsky's analysis of how cultural tools and signs shape human action~\cite{vygotsky1978mind}. Leontiev located individual actions within broader activities organized around motives and practical conditions~\cite{leont1978activity}. Engeström extended this account into a collective activity system that relates tool use to the social organization and outcome of an activity~\cite{engestrom1987learning}. This collective unit of analysis has been used in HCI to examine technology beyond isolated interface interaction~\cite{kuutti1992identifying,kaptelinin1997activity}.


For examining LLM-based family education, activity theory situates boundary coordination within a mediated collective activity~\cite{engestrom1987learning,kaptelinin1997activity}, extending Boundary Theory's focus on the coordination process itself~\cite{star1989boundary,akkerman2011boundary}. In LLM-based family education, such boundary processes may arise when parents and children hold different knowledge, expectations, or perspectives about an LLM-based artifact.
Further, activity theory extends the analysis from how coordination is achieved to how it is situated within and contributes to the educational object pursued through family interaction.
Once coordination is examined as part of an activity, the analysis also considers the conditions that organize it. Rules structure how participants may act, while the division of labor determines how educational responsibility is distributed. Community identifies the people and groups beyond the immediate interaction who share the educational object~\cite{engestrom1987learning}. This structure enables the review to analyze how LLM-based systems mediate the pursuit of educational objects. It also explains how rules, community relations, and the division of labor organize their use. Figure~\ref{fig:aodm-coding-components} presents these activity-system components and their relationships in the context of LLM-based family education.

The Activity-Oriented Design Method (AODM), developed by Mwanza, provides an operational bridge by translating the components and relations of an activity system into research questions and representations for empirical analysis~\cite{mwanza2002towards}. Prior applications have used AODM to analyze coordinated practices in healthcare~\cite{cornet2018activity,zellner2025addressing} and to examine learning activities in which participants use technologies to pursue shared educational objects~\cite{mwanza2009using,mwanza2011aodm}. In this review, AODM operationalizes activity theory across the corpus by structuring the analysis of how LLM-based systems mediate educational objects. It also enables comparison of how rules, communities, and divisions of labor organize participation.

\section{Scoping Review Method}
\label{Methodology}
This study used a scoping review method to map the range and characteristics of existing HCI research on LLM-based family education. The review covers studies of family use, family requirements for prospective systems, and LLM-supported analysis of family educational interactions. A scoping review~\cite{arksey2005scoping,tricco2018prisma} suited this mapping objective because the relevant research spans diverse study designs, technologies, family configurations, and educational settings. Accordingly, this approach allows the review to map the field's scope and diversity and identify which aspects of LLM-based family education have received limited research attention.

The review process was informed by Arksey and O'Malley's methodological framework for scoping studies~\cite{arksey2005scoping}. Reporting was guided by the PRISMA Extension for Scoping Reviews (PRISMA-ScR)~\cite{tricco2018prisma}. Arksey and O'Malley's framework structured the review stages. The research questions established the scope of inquiry (Section~\ref{sec:RQ}) and guided the development of the keyword-based search across selected HCI venues (Section~\ref{sec:search-strategy}). Records retrieved through the search underwent title and abstract screening, followed by full-text eligibility assessment using the exclusion criteria (Section~\ref{sec:study-selection}). PRISMA-ScR guided the transparent reporting of these stages and the subsequent data charting and synthesis. The selection process produced the final corpus for data charting and cross-study analysis (Section~\ref{sec:data-analysis}). At this stage, Activity Theory served as the analytical lens, while the Activity-Oriented Design Method (AODM) provided a common structure for charting and comparing the included studies. Using this structure, the review mapped how LLM-based systems mediate family educational activities and how family participation is organized around educational goals. In each findings section, the analysis first collates and summarizes patterns corresponding to the relevant research question, then identifies research gaps arising from limitations or fragmentation in the available evidence (Section~\ref{Findings}).

\subsection{Research Questions}\label{sec:RQ}

To structure the review's analytical scope, activity theory was operationalized through the Activity-Oriented Design Method (AODM), which represents an activity as a system of interrelated components~\cite{mwanza2002towards}. In this system, subjects pursue an object through mediating tools, and transforming the object produces outcomes. The activity is situated within a community, regulated by rules, and organized through a division of labor.

Building on the use of AODM in mobile learning research~\cite{mwanza2009using}, its activity-system components were adapted to LLM-based family education to formulate the following research questions (Figure~\ref{fig:aodm-coding-components}):

\begin{figure*}[t]
    \centering
    \includegraphics[width=0.98\textwidth]{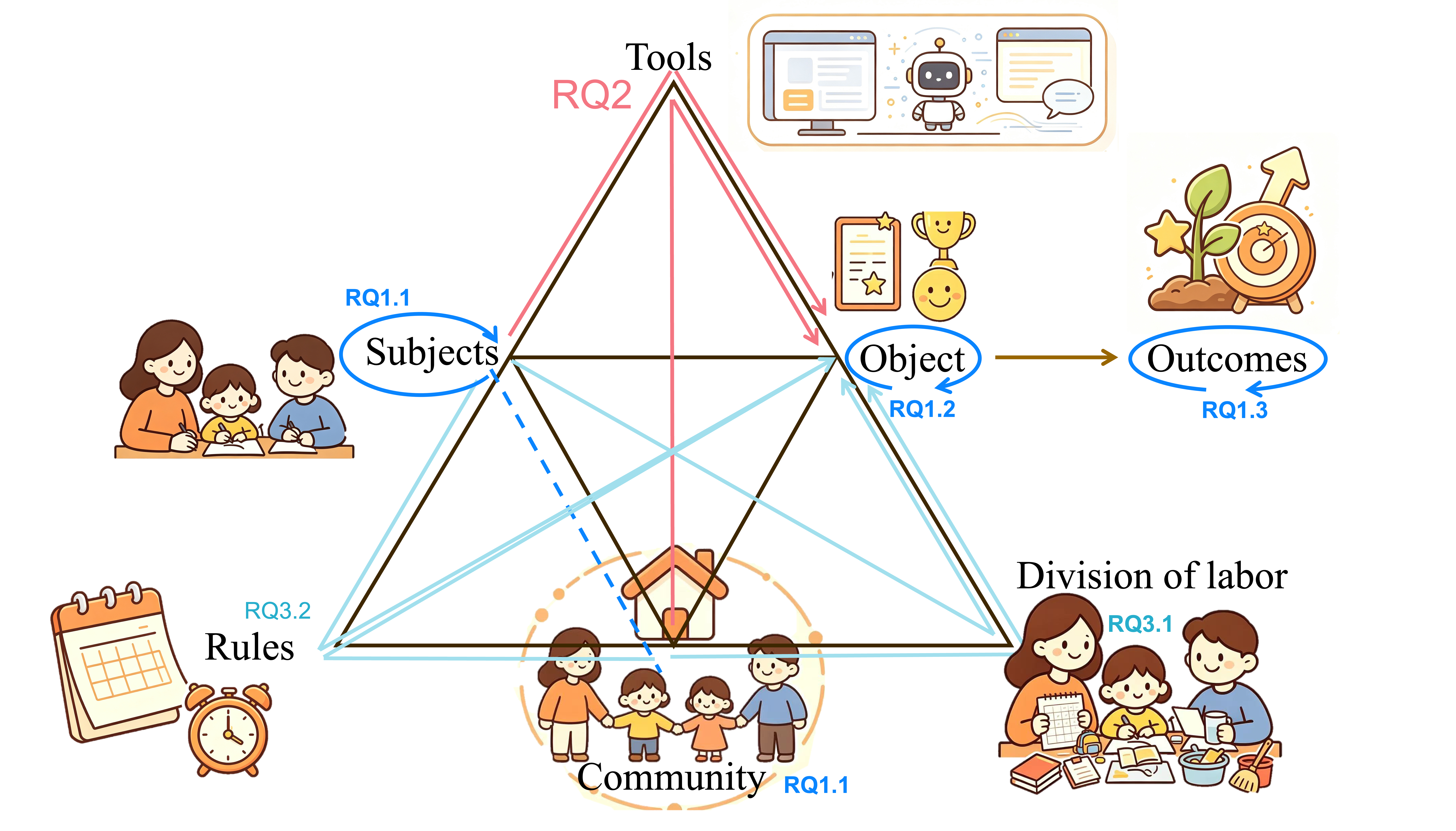}
    \caption{AODM-based analytical framework linking activity-system components to the three research questions.}
    \Description{An activity-system diagram retaining the triangular relations among tools, subjects, object, rules, community, and division of labor, with the object leading to outcomes. Icons and short descriptions explain each component, while labels map the components to RQ1, RQ2, and RQ3.}
    \label{fig:aodm-coding-components}
\end{figure*}

\begin{itemize}
    \item \textbf{RQ1: Family education activities.} \textbf{RQ1.1} identifies the family members and external communities that participate. \textbf{RQ1.2} examines the educational object that gives their participation a shared direction. \textbf{RQ1.3} identifies the outcomes produced through the activity and the evidence used to evaluate them.

    \item \textbf{RQ2: Technical capabilities and implementations of LLM-based systems.} What technical capabilities and system implementations enable LLM-based technologies to mediate the pursuit of educational objectives by subjects and communities?

    \item \textbf{RQ3: Organization of LLM use within family education activities.} Building on the activity context established in RQ1 and the technological analysis in RQ2, \textbf{RQ3.1} examines the division of labor associated with LLM use across the participants in an activity. \textbf{RQ3.2} then examines the rules through which participation and LLM use are organized.
\end{itemize}

\subsection{Search Strategy and Record Identification}\label{sec:search-strategy}
\begin{figure}[t]
    \centering
    \includegraphics[width=\columnwidth]{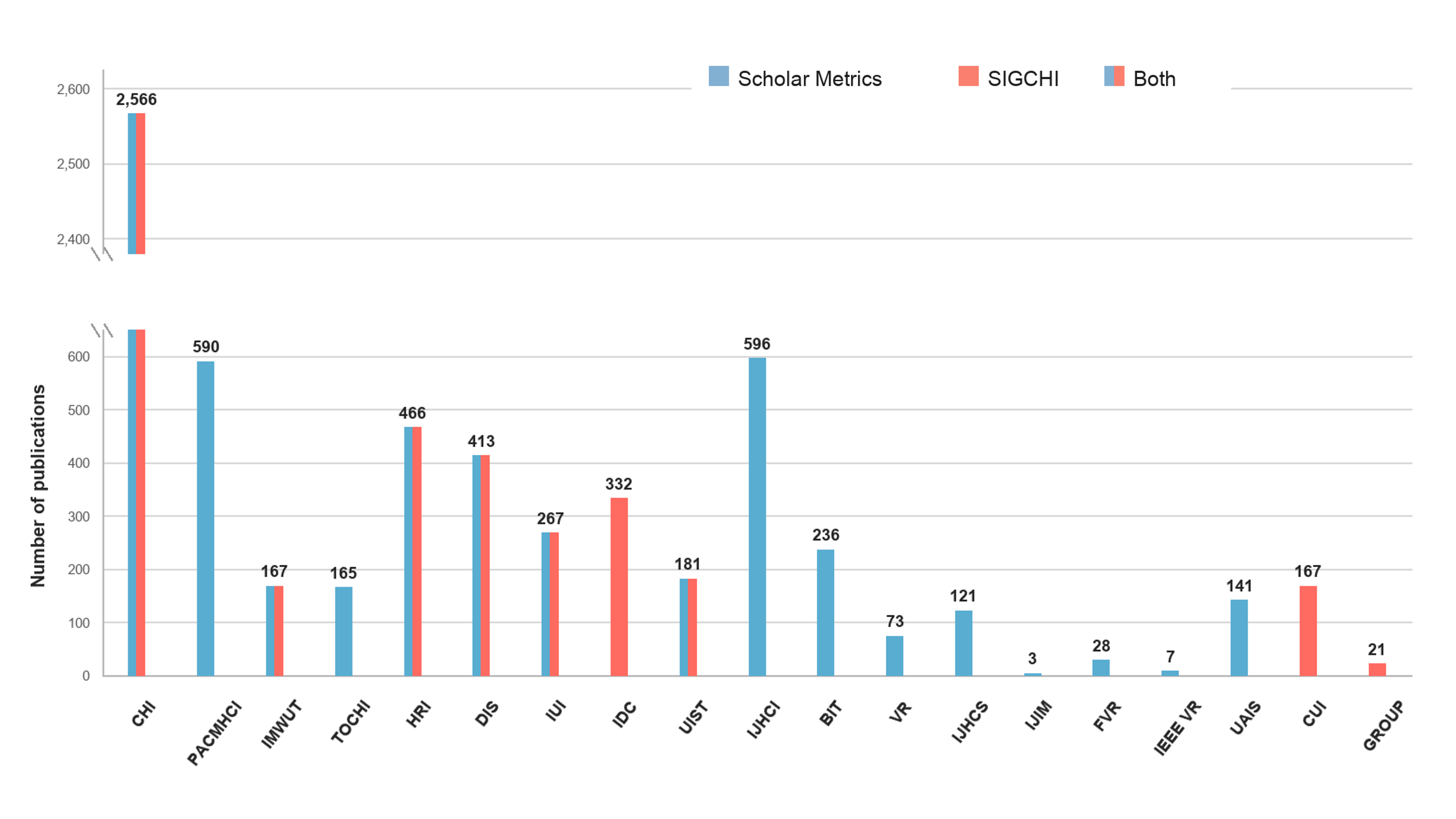}
    \caption{The distribution of publications across selected venues identified through Scholar Metrics and SIGCHI. Split-color bars indicate venues included in both sources.}
    \Description{Bar chart comparing publication counts across the searched HCI venues. Bar colors distinguish venues identified through Google Scholar Metrics, SIGCHI, or both sources.}
    \label{fig:venue_distribution}
\end{figure}

Venue-based sampling defined the source boundary of the review. The initial venue list comprised the top 20 venues in the Human Computer Interaction category of Google Scholar Metrics~\cite{googleScholarMetricsHCI} and was supplemented with conferences and journals associated with ACM SIGCHI~\cite{acmSIGCHI}. Each selected venue was searched separately. Figure~\ref{fig:venue_distribution} presents the number of records retrieved from each venue and indicates whether the venue was identified through Scholar Metrics, SIGCHI, or both sources.

The initial query centered on three concepts: LLM-based technology, education, and family context. To expand the query, the author team reviewed author-defined keywords and terminology in the titles and abstracts of seed papers on child- and parent-facing LLM systems in family education. We expanded the terms for LLM-based technologies because HCI papers may describe an application using terms such as \textit{chatbot}, \textit{conversational agent}, or \textit{agent} rather than explicitly identifying the underlying language model. We expanded the education terms to capture different descriptions of teaching and learning activities, and the family terms to cover both family participants and the settings in which family education occurs. We then checked the expanded query against the seed papers and refined it when it failed to retrieve a known relevant paper. In the final query, terms within each concept were combined with OR, and the three concepts were combined with AND:

\begin{quote}
\small
\texttt{
("LLM" OR "large language model" OR "generative AI" OR "chatbot" OR "conversational agent" OR "agent")
\textbf{AND} ("education" OR "learning" OR "pedagogy" OR "teaching")
\textbf{AND} ("family" OR "families" OR "parents" OR "home" OR "caregiver" OR "child" OR "children" OR "sibling" OR "grandparent")
}
\end{quote}

The same three concept groups and Boolean structure were used across the selected venues. The number of records retrieved from each venue was recorded separately and is presented in Figure~\ref{fig:venue_distribution}. The publisher platforms and search keywords are reported in Appendix 1.

No lower publication date bound was imposed. Although GPT-3 marked an important milestone in the development and visibility of large language models in 2020~\cite{brown2020language}, public attention to conversational generative AI accelerated substantially after the release of ChatGPT in November 2022~\cite{openai2022chatgpt}. Searching across all publication years reduced the risk of excluding conceptually relevant antecedent work.

We completed the final venue searches on May 4, 2026. The query was applied to publication metadata, including titles, abstracts, and author keywords, and identified 6,540 records across the selected venues. Covidence removed 83 duplicate records. Subsequent manual comparison of records with identical normalized titles identified two additional duplicate records; we retained their DOI-complete versions, leaving 6,455 unique records for title-and-abstract screening.

\subsection{Study Selection and Eligibility Assessment}\label{sec:study-selection}

\subsubsection{Criteria Development and Calibration}
The two authors jointly developed preliminary eligibility criteria based on the review scope and research questions. They then independently classified the same randomly sampled pilot set of 50 titles and abstracts as include or exclude using the preliminary criteria and compared their decisions to identify ambiguous definitions and inconsistent interpretations. Before consensus, the authors agreed on 49 of the 50 records (98\% raw agreement), with a binary Cohen's $\kappa$ of 0.960. These agreement statistics describe the 50-record calibration exercise, not the subsequent formal screening. The authors discussed the single disagreement and refined the criteria and accompanying decision rules before formal screening. These discussions produced the inclusion and exclusion criteria presented below, which were applied to the remaining records and used to reassess the pilot set.

A study was eligible when it satisfied all of the following inclusion criteria:

\begin{itemize}
    \item \textbf{IC 1: Primary empirical evidence.}
    The paper reported analyzable findings from a completed study involving human participants or their interaction data.

    \item \textbf{IC 2: Family-organized participation.}
    A family relationship shaped who participated, how participants interacted, or how learning responsibilities were distributed.

    \item \textbf{IC 3: Explicit educational objective.}
    Learning, knowledge acquisition, or skill development was an intended objective of the focal activity, not an inferred benefit.

    \item \textbf{IC 4: Substantive LLM involvement.}
    An LLM substantively mediated the family educational activity or the analysis of family educational interactions.
\end{itemize}

The corresponding exclusion criteria were applied in the following order:

\begin{itemize}
    \item \textbf{EC 1: Secondary research.}
    The paper primarily synthesized prior literature rather than reporting a primary study.

    \item \textbf{EC 2: No completed empirical evidence.}
    The paper did not report analyzable findings from a completed study involving human participants or their interaction data.

    \item \textbf{EC 3: No explicit educational objective.}
    Learning or skill development was absent or appeared only as a possible secondary benefit.

    \item \textbf{EC 4: No substantive LLM involvement.}
    The LLM was absent or used only incidentally, without contributing to the family educational activity or its analysis.

    \item \textbf{EC 5: Incidental family context.}
    This criterion was assessed last. Family relationships were mentioned but did not shape participation, interaction, or learning responsibilities.
\end{itemize}

When multiple criteria applied, we recorded the first applicable criterion as the primary exclusion reason.

\subsubsection{Screening Process and Results}
\textbf{Title-and-abstract screening.} Two researchers independently screened all 6,455 unique records. They retained 223 and 218 records, respectively, and agreed on the inclusion of 213 records. Their decisions showed 99.77\% raw agreement and a Cohen's $\kappa$ of 0.965. Disagreements were resolved through discussion, producing a consensus set of 213 reports for full-text assessment. We applied the exclusion criteria sequentially from EC1 to EC5 and excluded a record only when the available metadata clearly met the applicable criterion. Several boundary cases required additional decision rules: 1) Publication length or contribution type did not determine eligibility when a paper reported completed empirical evidence. 2) Studies could also qualify when an LLM was used by researchers to analyze family educational interactions rather than as a system used directly by family members~\cite{gao2025homework,kalanadhabhatta2024playlogue,shi2026towards}. 3) Eligible studies demonstrated substantive LLM involvement in educational interactions organized through family relationships. The screening excluded 6,242 records: 258 under EC1, 188 under EC2, 3,763 under EC3, 16 under EC4, and 2,017 under EC5.

\textbf{Full-text eligibility assessment.} All 213 reports that advanced from title-and-abstract screening were retrieved. Two researchers independently assessed the full texts and retained 53 and 54 studies, respectively. They agreed on 53 inclusions, corresponding to 99.53\% raw agreement and a Cohen's $\kappa$ of 0.988. The single disagreement was resolved through discussion. No report was excluded because the full text was unavailable. All retrieved reports were in English and were assessed against the complete eligibility criteria. Applying EC1--EC5 sequentially, the full-text assessment excluded 160 reports: 47 under EC1, 23 under EC2, 67 under EC3, 3 under EC4, and 20 under EC5.
 The remaining 53 studies satisfied IC1--IC4 and formed the final corpus.

\subsection{Data Analysis}\label{sec:data-analysis}

The analysis used AODM to define the activity components examined across the corpus. Inductive coding was then used to develop categories within those components. Each study served as the unit of analysis. We identified each study's focal activity configuration from its central research question, design goal, or empirical objective. This interpretation was checked against the system description, study design, findings, and discussion. When a study described multiple configurations, we treated the configuration most closely aligned with its central objective as the focal configuration. Other configurations were retained as contextual information. The object, subjects, tools, rules, community, division of labor, and outcomes were then recorded in a study-level coding record (Appendix 2, Tables~\ref{tab:appendix-activity-context} and~\ref{tab:appendix-activity-organization}).

Two researchers coded the same initial set of 10 studies to calibrate the framework. Their interpretations were largely aligned for most activity components. Rules required more extensive discussion because studies did not always distinguish system controls from expectations established by families or external institutions. The researchers used these discussions to clarify component definitions and formulate the decision rules in the codebook shown in Table~\ref{tab:coding-logic}. After calibration, the remaining 43 studies were divided among three researchers and coded using the resulting codebook. When discussion changed a component definition or category boundary, previously coded studies that could be affected were examined again.

We developed categories within each activity component by comparing the completed study-level records. For the object, the purpose organizing the focal configuration was coded as primary when a study reported multiple purposes. Additional purposes informed the interpretation of layered or overlapping categories but were not counted as separate cases. The resulting records were then compared across studies to identify how family educational activities were configured, how LLM-based systems were implemented, and how rules and divisions of labor organized their use. Categories produced through this analysis are reported with the corresponding findings.

The synthesis distinguishes systems used in family activity, support mediated by a family member or professional, and tools used by researchers to analyze interaction records. Research participation or representation in a dataset does not establish that a family used the LLM. Similarly, the reported-findings column in Appendix Table~\ref{tab:appendix-activity-context} includes learning or interaction findings, participant perceptions, design requirements, and technical results. These forms of evidence answer different questions: requirements describe desired support, and computational accuracy describes an analysis tool; neither alone demonstrates educational change.

\newcommand{\AODMCodebook}{%
\begingroup
\small
\renewcommand{\arraystretch}{1.18}
\begin{xltabular}{\textwidth}{p{2.6cm} X}
\caption{AODM-based coding framework for LLM-based family education activities.}
\label{tab:coding-logic} \\
\toprule
\textbf{Activity component} & \textbf{Operational coding rule} \\
\midrule
\endfirsthead
\multicolumn{2}{l}{\small\textit{Table~\thetable\ continued from the previous page}} \\
\toprule
\textbf{Activity component} & \textbf{Operational coding rule} \\
\midrule
\endhead
\midrule
\multicolumn{2}{r}{\small\textit{Continued on the next page}} \\
\endfoot
\bottomrule
\endlastfoot

Object
& The primary purpose or problem that organizes the focal activity. The object pursued is distinguished from the results produced through the activity~\cite{engestrom2014learning,mwanza2002towards}. \\

Subjects 
& Actors whose actions are directed toward the object. For needs studies and retrospective analyses, distinguish research participants or actors represented in records from actual users of an LLM-based system. \\

Tools 
& The mediation performed by the LLM-based system and its interface or resources. Distinguish use within family activity from researcher use to analyze that activity. Distinct system functions may receive multiple tool codes. \\

Rules 
& Explicit controls, protocols, expectations, norms, or design requirements that regulate who may act, which system contributions may enter the activity, and how information may move. \\

Community 
& Groups beyond the focal subjects that share, influence, or hold a stake in the object. Inclusion requires evidence of a substantive relation to the activity and is not inferred from the setting alone. \\

Division of labor 
& The distribution of tasks, responsibility, and authority among the system, family members, and community participants. The analysis records work delegated to the system alongside work retained by people. \\

Outcomes 
& Changes in learning or interaction supported by the reported evidence. The appendix also summarizes perceptions, design requirements, and technical findings; these are identified as such where relevant and do not, by themselves, establish an educational outcome. \\

\end{xltabular}
\endgroup
}

\section{Findings}

\label{Findings}

The findings distinguish reported use from prospective designs and retrospective analysis. They first describe the reviewed corpus and then follow the activity-system relations represented in the research questions. Findings 1 establishes the subjects, communities, educational objects, and reported outcomes of the activities. Findings 2 explains how LLM capabilities and system implementations mediated those objects. Findings 3 examines how this mediation distributed labor and was regulated by rules. Each part concludes by identifying a limitation in the evidence available for explaining the corresponding activity-system relation.

\subsection{Corpus Overview}\label{sec:findings-overall}
The final corpus consists of 53 studies selected from an initial pool of 6,540 records retrieved across 19 HCI venues associated with SIGCHI, listed in Google Scholar Metrics, or identified by both sources (Sections~\ref{sec:search-strategy} and~\ref{sec:study-selection}). Publication increased sharply after 2023: the corpus includes one study from 2022, one from 2023, 10 from 2024, 23 from 2025, and 18 from 2026. This concentration follows the accelerated adoption of generative AI after the release of ChatGPT in November 2022~\cite{openai2022chatgpt}; only one study was published before 2023.

The 53 studies appeared in 10 venues. CHI accounted for 30 studies, followed by IDC with nine and PACMHCI with four; the remaining 10 studies were distributed across seven venues. The corpus includes studies that designed and evaluated interactive systems, qualitative investigations of family needs and practices, design explorations, and work centered on datasets or computational assessment. These forms can overlap within a single study and were treated as methodological characteristics instead of mutually exclusive contribution categories during data charting (Section~\ref{sec:data-analysis}).

\subsection{Findings 1: Participants, Educational Objects, and Outcomes in LLM-Based Family Education}\label{Findings-RQ1}
\subsubsection{Child--Parent Relationships and Selective External Expertise (RQ1.1)}

The reviewed studies most often addressed children in the kindergarten and elementary-school age bands used in Table~\ref{tab:target-child-audience-citations}: 29 studies addressed ages 5--8 and 25 addressed ages 7--11. The corresponding counts were 13 for ages 3--4, 14 for ages 12--14, and eight for ages 15--18. These counts describe the ages addressed by each study, including intended users in research conducted with parents. The age bands overlap, and a study can appear in several rows.

\begin{table}[H]
\centering
\small
\caption{Child age groups addressed in the reviewed studies. Age bands and study counts overlap; rows include intended users in parent-focused research.}
\label{tab:target-child-audience-citations}
\begin{tabular}{p{2.2cm}p{2.2cm}p{1.2cm}p{7.5cm}}
\toprule
Age group & Age range & Studies & References \\
\midrule
Preschool & Ages 3--4 & 13 &
\cite{chen2025characterizing,sun2024exploring,kalanadhabhatta2024playlogue,ho2025set,dangol2025want,chheda2025artinsight,liu2024he,lee2024open,dietz2024contextq,he2025storypal,shi2026towards,zhang2026parents,rifat2026homeroom} \\

Kindergarten & Ages 5--8 & 29 &
\cite{chen2025characterizing,liu2025bricksmart,zhao2025youthcare,shen2025easel,sun2024exploring,gao2025homework,kalanadhabhatta2024playlogue,ho2025set,wang2025charactercritique,olutunbi2026leads,long2022family,dangol2025want,chheda2025artinsight,han2024teachers,liu2024he,lee2024open,kim2025bridging,dietz2024contextq,he2025storypal,liu2026dollama,shi2026towards,viswanathan2025interaction,xu2025accompany,seo2025enhancing,driscoll2026understanding,xie2026understanding,nawshin2026well,zhang2026parents,rifat2026homeroom} \\

Elementary school & Ages 7--11 & 25 &
\cite{chen2025characterizing,liu2025bricksmart,zhao2025youthcare,shen2025easel,gao2025homework,figueiredo2025designing,olutunbi2026leads,long2022family,dangol2025want,chheda2025artinsight,han2024teachers,liu2024he,kim2025bridging,yang2026autiverse,dietz2024contextq,he2025storypal,liu2026dollama,shi2026towards,xu2025accompany,seo2025enhancing,driscoll2026understanding,xie2026understanding,nawshin2026well,zhang2026parents,rifat2026homeroom} \\

Middle school & Ages 12--14 & 14 &
\cite{zhao2025youthcare,long2022family,dangol2025want,chheda2025artinsight,han2024teachers,kim2025bridging,yang2026autiverse,xu2025accompany,seo2025enhancing,driscoll2026understanding,xie2026understanding,nawshin2026well,zhang2026parents,rifat2026homeroom} \\

High school & Ages 15--18 & 8 &
\cite{zhao2025youthcare,long2022family,chheda2025artinsight,yang2026autiverse,driscoll2026understanding,zhang2025parental,zhang2026parents,rifat2026homeroom} \\

\bottomrule
\end{tabular}
\end{table}

Child--parent configurations dominated the corpus, accounting for 44 of the 53 studies (83\%; Table~\ref{tab:family-subject-configurations}). Configurations were defined by the family relationship that organized the focal activity. Child--parent studies included joint activities such as shared reading~\cite{dietz2024contextq,he2025storypal,wang2025charactercritique}, play~\cite{liu2025bricksmart}, and communication or media use~\cite{zhao2025youthcare}. They also included parent-only studies when adults' participation was directed toward a particular child's learning or development~\cite{sun2024exploring}, moderation of the child's interactions with generative AI~\cite{driscoll2026understanding}, or judgment of the child's readiness for independent technology use~\cite{xie2026understanding}. By comparison, parent-focused studies centered on adults' own learning or wellbeing without organizing the activity around a particular child--parent interaction~\cite{viswanathan2025interaction,petsolari2024socio}. The remaining nine studies comprised these two parent-focused studies, two child-focused configurations without a specified family co-subject~\cite{yang2026autiverse,olutunbi2026leads}, one grandparent--grandchild configuration~\cite{kim2025bridging}, and four adult-family or family--institutional arrangements~\cite{moon2026promises,wen2025families,kaur2025familyplanning,wu2026warmsteward}; no study centered a child--sibling configuration.

\begin{table}[H]
\centering
\small
\caption{Focal family relationships represented across the reviewed studies. Child--parent configurations include parent-only investigations about a child; the count does not imply that both participated.}
\label{tab:family-subject-configurations}
\renewcommand{\arraystretch}{1.15}
\begin{tabularx}{\textwidth}{@{}p{3.1cm}>{\centering\arraybackslash}p{1.2cm}X@{}}
\toprule
\textbf{Subject configuration} & \textbf{Number} & \textbf{Studies} \\
\midrule

Child-focused &
2 &
\cite{yang2026autiverse,olutunbi2026leads} \\
\addlinespace[2pt]

Child--parent &
44 &
\cite{wang2025charactercritique,chen2025characterizing,choi2025aacesstalk,liu2025bricksmart,zhao2025youthcare,shen2025easel,sun2024exploring,gao2025homework,figueiredo2025designing,kalanadhabhatta2024playlogue,ho2025set,long2022family,dangol2025want,chheda2025artinsight,han2024teachers,liu2024he,lee2024open,dietz2024contextq,he2025storypal,liu2026dollama,shi2026towards,xu2025accompany,seo2025enhancing,driscoll2026understanding,xie2026understanding,nawshin2026well,zhang2025parental,zhang2026parents,zaidi2025sociotechnical,wester2024facing,talai2025towards,rifat2026homeroom,seo2024chacha,yoo2026creativecollaboration,shi2025ineedyourhelp,song2026ambiguousloss,schiavo2026brainrot,zanardi2026dinner,antony2026ella,eira2025parents,han2023design,hu2024grow,hassan2026anyone,baba2026digitalinequality} \\
\addlinespace[2pt]

Child--grandparent &
1 &
\cite{kim2025bridging} \\
\addlinespace[2pt]

Child--sibling &
0 &
--- \\
\addlinespace[2pt]

Parent-focused &
2 &
\cite{viswanathan2025interaction,petsolari2024socio} \\
\addlinespace[2pt]

Other family arrangements &
4 &
\cite{moon2026promises,wen2025families,kaur2025familyplanning,wu2026warmsteward} \\

\bottomrule
\end{tabularx}
\end{table}

External communities entered family activity when knowledge or authority beyond the household was required (Table~\ref{tab:community-configurations}). The three community categories were distinguished by the kind of expertise they contributed to the focal activity. \textbf{Educational communities} included teachers who connected household activities with classroom expectations~\cite{chen2025characterizing,han2024teachers}. IEP stakeholders supplied institutional knowledge about special education processes~\cite{zaidi2025sociotechnical}. Homeschooling networks connected home learning with curriculum standards and trusted educational resources~\cite{rifat2026homeroom}. These contributions linked family activity with formal learning goals, curriculum expectations, and institutional procedures. \textbf{Clinical and care communities} included speech-language pathologists whose expertise informed home communication practice and the interpretation of parent--child joint attention~\cite{dangol2025want,shi2026towards}. Therapeutic and autism-support communities informed family interaction, communication scaffolding, and the interpretation of autistic adolescents' narratives~\cite{liu2024he,choi2025aacesstalk,yang2026autiverse}. Pediatric clinicians and accessibility experts supported healthcare communication and accessible artwork interpretation~\cite{seo2025enhancing,chheda2025artinsight}. These contributions addressed communication, development, therapy, health, disability, and accessibility. \textbf{Research and parenting communities} included research communities that supplied annotation practices for analyzing adult--child conversation~\cite{kalanadhabhatta2024playlogue}. Informal learning communities supported family discussion of AI literacy~\cite{long2022family}. Parenting-support communities informed acceptable forms of assistance for caregivers~\cite{viswanathan2025interaction,petsolari2024socio}. These contributions provided analytic practices, informal learning resources, and guidance for caregivers. Most remaining studies bounded the community at the family or household level.

\begin{table}[H]
\centering
\small
\caption{External communities represented in LLM-based family education activities.}
\label{tab:community-configurations}
\begin{tabularx}{\textwidth}{p{3.4cm}X p{3.7cm}}
\toprule
\textbf{Community type} & \textbf{Communities represented} & \textbf{Studies} \\
\midrule
Educational communities &
Teachers, education experts, IEP stakeholders, schools, and homeschooling networks. &
\cite{chen2025characterizing,han2024teachers,zaidi2025sociotechnical,rifat2026homeroom} \\

Clinical and care communities &
Speech-language pathologists, therapists, pediatric clinicians, healthcare providers, autism-support communities, and accessibility experts. &
\cite{choi2025aacesstalk,dangol2025want,liu2024he,yang2026autiverse,shi2026towards,seo2025enhancing,zhang2026parents,talai2025towards,wester2024facing,chheda2025artinsight} \\

Research and parenting communities &
Research communities, informal AI-learning communities, and parenting-support communities. &
\cite{kalanadhabhatta2024playlogue,long2022family,viswanathan2025interaction,petsolari2024socio} \\
\bottomrule
\end{tabularx}
\end{table}

\paragraph{Research gaps for RQ1.1.}
\textit{RG1.1a: Family relationships beyond children and parents remain underrepresented.} Forty-four studies addressed child--parent configurations; one centered grandparents and grandchildren~\cite{kim2025bridging}, and none centered siblings. Wider household relationships shape educational opportunities~\cite{marjoribanks2005family}: parents and siblings contribute different forms of teaching~\cite{fuoco2024parent}, and siblings learn through everyday family practice~\cite{maynard2002cultural,davies2019sticky}. The current concentration leaves limited evidence about how other family members shape guidance and authority around an LLM. Comparing similar activities across sibling, multigenerational, and multiple-caregiver configurations could reveal who initiates interaction, interprets outputs, and guides learning.

\textit{RG1.1b: The diversity of family circumstances remains insufficiently represented.} The corpus includes geographically separated families~\cite{xu2025accompany}, families with disability and communication needs~\cite{choi2025aacesstalk}, homeschooling families~\cite{rifat2026homeroom}, and families facing resource constraints~\cite{kaur2025familyplanning}. These studies demonstrate that LLM-based family education occurs under different relational, educational, and material circumstances. However, each circumstance was examined in a limited part of the corpus, while demographic characteristics and family resources were reported inconsistently. This limited coverage makes it difficult to determine whose experiences the current findings represent and whether the identified forms of participation extend across families. Parental involvement depends on how parents understand their responsibilities and whether they perceive themselves as capable of helping~\cite{hoover1997parents}. Access to resources further conditions how families from nondominant groups facilitate technology-based learning~\cite{yu2021parental}. Broader sampling would bring more family circumstances into view. Consistent reporting of family structure, caregiver roles, demographic characteristics, and access to educational resources would also show how these circumstances shape participation in LLM-supported education.

\subsubsection{Educational Objects Concentrated on Language, Interpretation, and Family Support (RQ1.2)}

The coding distinguished six educational object categories (Table~\ref{tab:content_category}). These categories describe intended learning purposes or the focus of a needs or interaction-analysis study; they do not indicate demonstrated learning gains. \textbf{Language, literacy, and narrative development} centered on vocabulary learning and personalized story reading~\cite{chen2025characterizing,lee2024open,antony2026ella}. Studies targeted dialogic participation during co-reading~\cite{dietz2024contextq,he2025storypal} and expressive communication through conversation or storytelling~\cite{choi2025aacesstalk,liu2024he}. Shared narratives helped grandparents and grandchildren learn about one another~\cite{kim2025bridging}, while stories connected geographically separated families~\cite{xu2025accompany}. Image-based storytelling supported refugee families in sharing memories and interpretations of home~\cite{song2026ambiguousloss}. \textbf{Logical and problem-solving development} focused on spatial relations and problem-solving methods during family block play. Parent-guided language connected these concepts with children's manipulation of physical materials~\cite{liu2025bricksmart}. \textbf{Cognitive development} appeared in activities that required participants to compare perspectives and critique story content~\cite{wang2025charactercritique}. Other studies examined observation and interpretation through parent--child joint-attention analysis~\cite{shi2026towards} or discussion of artwork among family members with different visual abilities~\cite{chheda2025artinsight}. \textbf{Emotional and relational development} was pursued through activities that helped children identify emotions and discuss them with parents~\cite{shen2025easel,seo2024chacha}. Shared storymaking enabled children to express experience with parents and therapists~\cite{liu2024he}. Other studies addressed communication and mutual understanding across generations~\cite{kim2025bridging}, geographic distance~\cite{xu2025accompany}, healthcare encounters~\cite{seo2025enhancing,wester2024facing}, and developmental support contexts~\cite{talai2025towards}. \textbf{AI and digital literacy development} made the technology itself a learning object. Families examined how generative AI works and how to use it responsibly~\cite{long2022family,wu2026warmsteward}. Parents and children evaluated chatbot use~\cite{driscoll2026understanding}, video appropriateness~\cite{zhao2025youthcare,nawshin2026well}, and trust in AI~\cite{zhang2025parental}. Parents further considered how AI use aligned with educational expectations~\cite{zaidi2025sociotechnical}, family values~\cite{petsolari2024socio,rifat2026homeroom}, and children's support needs~\cite{zhang2026parents}. \textbf{Self-directed learning and learning habits} concerned children's growing independence in learning with generative AI. Parents described children's readiness and approaches to mediating increasingly independent AI use~\cite{xie2026understanding}.

\begin{table}[H]
\centering
\small
\caption{Educational object categories and examples from the reviewed studies.}
\label{tab:content_category}
\begin{tabularx}{\textwidth}{p{3.4cm}X p{3.6cm}}
\toprule
\textbf{Educational object} & \textbf{Examples in the corpus} & \textbf{References} \\
\midrule
Language, Literacy, and Narrative Development &
Personalized vocabulary and story reading, generated questions for parent--child co-reading, conversational support, collaborative storymaking, visual narratives, and intergenerational family storytelling. &
\cite{chen2025characterizing,sun2024exploring,kalanadhabhatta2024playlogue,wang2025charactercritique,lee2024open,dietz2024contextq,kim2025bridging,liu2024he,yang2026autiverse,olutunbi2026leads,dangol2025want,he2025storypal,han2024teachers,figueiredo2025designing,liu2026dollama,ho2025set} \\

Logical and Problem-Solving Development &
Parent-guided spatial language and problem solving during family block play. &
\cite{liu2025bricksmart} \\

Cognitive Development &
Critical comparison of story perspectives, observation and interpretation of parent--child joint attention, and accessible interpretation of artwork. &
\cite{wang2025charactercritique,shi2026towards,chheda2025artinsight} \\

Emotional and Relational Development &
Emotional reflection during media use, family storymaking, interpretation of homework interaction, communication across distance or generations, pediatric communication, and developmental guidance. &
\cite{viswanathan2025interaction,shen2025easel,kim2025bridging,xu2025accompany,gao2025homework,seo2025enhancing,liu2026dollama,talai2025towards,wester2024facing} \\

AI and Digital Literacy Development &
Family discussion of AI concepts, parental moderation of chatbot use, assessment of video appropriateness, responsible GenAI use, and alignment of AI support with educational expectations and family values. &
\cite{long2022family,driscoll2026understanding,xie2026understanding,nawshin2026well,zhang2025parental,zhao2025youthcare,petsolari2024socio,zaidi2025sociotechnical,zhang2026parents,rifat2026homeroom} \\

Self-Directed Learning and Learning Habits &
Parents' perspectives on readiness for independent generative AI use and on appropriate mediation. &
\cite{xie2026understanding} \\
\bottomrule
\end{tabularx}
\end{table}

\paragraph{Research gaps for RQ1.2.}
\textit{RG1.2a: Understanding how LLMs mediate diverse forms of family learning requires a broader evidence base.} Personalized story reading represented language learning through generated narratives~\cite{chen2025characterizing}. ContextQ examined questions for parent--child co-reading~\cite{dietz2024contextq}, while StoryPal used LLM-generated prompts to support dialogic reading~\cite{he2025storypal}. Other studies used generated stories to support emotional reflection~\cite{seo2024chacha} or examined family judgments of generated media content~\cite{schiavo2026brainrot}. Spatial-language learning during block play was one of the few activities that connected generated guidance with physical problem solving~\cite{liu2025bricksmart}. This distribution leaves mathematical reasoning, scientific inquiry, and embodied learning weakly represented. Home learning research connects family activity with literacy development~\cite{melhuish2008effects} and numeracy development~\cite{soto2020identifying}. Informal learning research identifies families and everyday environments as settings for science inquiry~\cite{national2009learning}. Formal teaching and informal shared activities also contribute differently to children's early learning~\cite{skwarchuk2014formal}. A broader evidence base can explain how LLMs mediate educational objects across intentional instruction and learning embedded in family activity. Future research should examine LLM-supported mathematical, scientific, and embodied activities and compare how learning is organized across formal and informal family settings.

\textit{RG1.2b: Explaining how educational objects develop through family routines requires longitudinal evidence.} Several reviewed activities were organized around bounded sessions. ContextQ examined questions generated for a co-reading session~\cite{dietz2024contextq}, while StoryPal examined dialogic reading with young children~\cite{he2025storypal}. Storymaking also provided a focal activity in family expressive arts therapy~\cite{liu2024he}; Culture Link examined family storytelling over a ten-day field trial~\cite{song2026ambiguousloss}. Media-oriented studies examined collaborative content evaluation~\cite{zhao2025youthcare} or judgments of video appropriateness~\cite{nawshin2026well}. Research on children's self-directed learning addressed changing parental involvement~\cite{xie2026understanding}, and Homeroom situated curriculum development within homeschooling practice~\cite{rifat2026homeroom}. These studies begin to connect LLM use with ongoing family organization. A longitudinal account is important because planning, monitoring, and self-regulation develop through repeated practice~\cite{zimmerman2002becoming}. Learners' interests also acquire educational significance when they are sustained across activities and settings. Parents adjust home-learning activities in response to children's changing performance~\cite{silinskas2020responsive}. Longitudinal HCI research can trace how educational objects, children's development, and family support shape one another across recurring homework, study, caregiving, and intergenerational learning routines.

\subsubsection{Reported Outcomes and Their Evidence (RQ1.3)}

Table~\ref{tab:outcome-evidence-distinction} groups the reported outcomes and related findings by their educational focus. The studies provide different kinds of evidence, so these groups should not be read as four demonstrated classes of learning gain.

\textbf{Children's learning and development.} Evaluations examined spatial language during family block play~\cite{liu2025bricksmart}, vocabulary assessment and learning~\cite{lee2024open,antony2026ella}, critical interpretation of stories~\cite{wang2025charactercritique}, emotional identification~\cite{seo2024chacha}, narrative expression~\cite{yang2026autiverse}, and engagement during dialogic reading~\cite{he2025storypal}. These measures concern different aspects of participation and development. By comparison, the self-directed learning study elicited parents' perspectives on readiness and mediation; it did not demonstrate an increase in children's readiness~\cite{xie2026understanding}.

\textbf{Parental learning and support.} Studies examined communication strategies in parent--child interaction~\cite{choi2025aacesstalk}, parents' requirements for home speech practice~\cite{dangol2025want}, support for navigating special-education processes~\cite{zaidi2025sociotechnical}, and intergenerational learning about responsible GenAI use~\cite{wu2026warmsteward}. Accounts of desired or co-designed support establish what parents need, whereas claims of improved parental knowledge require evidence of learning.

\textbf{Shared family learning and interaction.} Studies examined dialogue and turn-taking~\cite{choi2025aacesstalk,dietz2024contextq}, shared media reflection~\cite{shen2025easel}, and intergenerational storytelling~\cite{kim2025bridging,xu2025accompany,song2026ambiguousloss}. Other work elicited home-practice requirements~\cite{dangol2025want} or combined family design sessions with expert appraisal of a pediatric chatbot~\cite{seo2025enhancing}. Homeroom examined how generated materials fitted homeschooling practice~\cite{rifat2026homeroom}. These findings concern different stages of design and use; they do not establish lasting educational effects.

\textbf{Judgments and expectations about LLM use.} Studies elicited media-appropriateness judgments~\cite{zhao2025youthcare,nawshin2026well}, moderation preferences~\cite{driscoll2026understanding}, views on autonomy~\cite{xie2026understanding}, expectations for household agents~\cite{wen2025families}, and values expressed through discussion of generated content~\cite{zanardi2026dinner}. Trust and acceptance measures describe participants' perceptions of AI and proposed uses~\cite{zhang2025parental,zhang2026parents}. These findings inform design and adoption; they are not interchangeable with evidence that families learned to use LLMs more effectively.

\begin{table}[H]
\centering
\small
\caption{Reported outcomes and related findings, with limits on their interpretation.}
\label{tab:outcome-evidence-distinction}
\begin{tabularx}{\textwidth}{p{2.7cm}X p{3.7cm}p{4.0cm}}
\toprule
\textbf{Educational focus} & \textbf{Reported outcomes and findings} & \textbf{Evidence and interpretation} & \textbf{References} \\
\midrule
Children's learning and development & Language, expression, engagement, and parental views on readiness & Learning and task measures, observations, logs, and interviews; views on readiness do not establish increased readiness & \cite{liu2025bricksmart,lee2024open,antony2026ella,wang2025charactercritique,seo2024chacha,yang2026autiverse,he2025storypal,xie2026understanding} \\
Parental learning and support & Communication strategies, support needs, and approaches to educational guidance & Observations, interviews, surveys, and parent reports; requirements alone do not demonstrate learning & \cite{choi2025aacesstalk,dangol2025want,zaidi2025sociotechnical,wu2026warmsteward} \\
Shared family learning and interaction & Dialogue, participation, coordination, and shared interpretation & Conversation measures, observations, family reports, and expert appraisal; design feedback is distinct from observed interaction and sustained effects & \cite{choi2025aacesstalk,dietz2024contextq,shen2025easel,kim2025bridging,xu2025accompany,song2026ambiguousloss,dangol2025want,seo2025enhancing,rifat2026homeroom} \\
Judgments about LLM use & Content assessments, moderation preferences, trust, acceptance, and values & Judgment tasks, surveys, and elicited preferences; perceptions and intentions are not learning gains & \cite{zhao2025youthcare,nawshin2026well,driscoll2026understanding,xie2026understanding,wen2025families,zanardi2026dinner,zhang2025parental,zhang2026parents} \\
\bottomrule
\end{tabularx}
\end{table}

\paragraph{Research gaps for RQ1.3.}
\textit{RG1.3a. Evaluations rarely connect individual learning, family interaction, and technical performance.} Existing studies established different parts of this relationship. AACessTalk documented changes in turn-taking and communication between children and parents~\cite{choi2025aacesstalk}. eaSEL examined shared reflection during media use~\cite{shen2025easel}. Intergenerational co-creation research examined connection between grandparents and grandchildren~\cite{kim2025bridging}. Other contributions stopped at identifying parents' requirements for children's storytelling systems~\cite{sun2024exploring}, expectations for AI parenting support~\cite{petsolari2024socio}, or technical performance in analyzing adult--child conversations~\cite{kalanadhabhatta2024playlogue}. This separation prevents the corpus from showing whether technical capability or acceptability improves learning through changes in family participation. Reviews of parent--child joint media engagement indicate that children's outcomes depend partly on the interaction organized around media use~\cite{ewin2021impact,yu2024joint}. Evaluating system performance alongside changes in family interaction and educational outcomes within the same study would make this relationship visible.

\subsection{Findings 2: How LLM-Based Systems Mediate Family Education}
\label{Findings-RQ2}

For family-facing systems, LLM capabilities became educationally relevant through their interfaces and participant arrangements. In retrospective research, LLMs instead helped analyze records of interaction. Analytic performance is considered separately from evidence that a system mediated family participation. Four forms of mediation recurred across the corpus: producing materials, structuring dialogue, interpreting contextual information, and connecting family activity with external knowledge and values. A system could combine several forms, while the same capability could assume different educational roles as family participation changed.

\subsubsection{LLM Contributions Across Interface Forms}

Table~\ref{tab:prellm-llm-comparison} relates interface forms to the contributions described in the reviewed systems. The comparison concerns how support is arranged; it does not estimate gains over earlier technologies.

\begin{table}[H]
\centering
\small
\caption{Interface forms and generative contributions described in the reviewed systems. The rows summarize system roles rather than comparative evidence of improvement.}
\label{tab:prellm-llm-comparison}
\renewcommand{\arraystretch}{1.15}
\begin{tabularx}{\textwidth}{@{}>{\raggedright\arraybackslash}p{2.1cm}>{\raggedright\arraybackslash}p{3.2cm}>{\raggedright\arraybackslash}p{4.2cm}>{\raggedright\arraybackslash}X@{}}
\toprule
\textbf{System form} & \textbf{Interface contribution} & \textbf{Generative contribution described} & \textbf{Examples in the corpus} \\
\midrule
Conversational systems & Organizes an exchange through questions, responses, and turn-taking & Generates follow-up questions, prompts, or explanations using preceding turns and participant input & Co-reading questions~\cite{dietz2024contextq,he2025storypal}. Multi-agent critique~\cite{wang2025charactercritique}. Contextual parent guidance~\cite{choi2025aacesstalk}. Emotion elicitation~\cite{seo2024chacha}. Multimodal journaling~\cite{yang2026autiverse}. \\
\addlinespace[3pt]
Physically embodied systems & Uses physical presence, gaze, gesture, and turn-taking to coordinate attention & Generates stories and dialogue around learning targets, with adult input or supervision where provided & Parent-adjustable educational robots~\cite{ho2025set}. Autonomous and adult-supported storytelling~\cite{olutunbi2026leads}. Language-learning robots~\cite{antony2026ella}. Robots mediating family information~\cite{wester2024facing}. \\
\addlinespace[3pt]
Spatial or immersive systems & Connects interaction with physical objects, visual artifacts, or spatial representations & Uses physical constructions, toys, drawings, or images as context for generated guidance and interpretation & Family block play~\cite{liu2025bricksmart}. Multimaterial storymaking~\cite{liu2024he}. Toy-mediated drama~\cite{liu2026dollama}. AI-supported sandplay~\cite{shi2025ineedyourhelp}. \\
\bottomrule
\end{tabularx}
\end{table}

\textbf{Conversational interfaces organize an exchange.} Earlier studies showed how families adapted speech and repaired breakdowns around household conversational agents~\cite{beneteau2019communication,sciuto2018hey,xu2020exploring,du2021alexa}. In the reviewed LLM-based systems, preceding turns and participant input informed generated questions, responses, or guidance. ContextQ and StoryPal supported parent--child co-reading~\cite{dietz2024contextq,he2025storypal}; CharacterCritique supplied contrasting perspectives on stories~\cite{wang2025charactercritique}; and AACessTalk generated prompts for parent--child communication~\cite{choi2025aacesstalk}. These contributions place generated material within an unfolding exchange. Their educational role depends on how participants interpret it and continue the interaction.

\textbf{Physical embodiment coordinates attention and participation.} Earlier robots used presence, gaze, gesture, and turn-taking in children's activities~\cite{scassellati2018teaching,cagiltay2020investigating,ho2024s}. The reviewed systems combined these affordances with generated content. SET-PAiREd included parental review and adjustable involvement~\cite{ho2025set}. Storytelling robots operated with different degrees of adult support~\cite{olutunbi2026leads}, and ELLA generated dialogue around parent-selected vocabulary~\cite{antony2026ella}. The relevant distinction is how generation and adult participation are arranged within the embodied interaction, rather than whether an earlier robot could adapt.

\textbf{Material and spatial interfaces connect support with situated activity.} Earlier systems linked physical objects with representations of circuits or mathematical attributes~\cite{beheshti2017wires,kang2020armath}. BrickSmart used family block play to inform spatial-language guidance~\cite{liu2025bricksmart}. Other reviewed designs combined generated material with family storymaking~\cite{liu2024he}, toy-mediated drama~\cite{liu2026dollama}, or interpretation of sandplay artifacts~\cite{shi2025ineedyourhelp}. These cases show how a physical artifact can become context for a generated contribution. They do not establish that generation improves on other ways of supporting the same activity.

\subsubsection{Mediation Functions and Educational Objects}

Table~\ref{tab:forms-of-llm-mediation} summarizes functions described or proposed for family activities, including applications whose evaluations examined requirements or analytic performance.

\begin{table}[H]
\centering
\small
\caption{Recurring functions of systems in the reviewed family activities, including generative media and research analysis tools. Technical components and evaluation evidence differ across examples.}
\label{tab:forms-of-llm-mediation}
\begin{tabularx}{\textwidth}{p{2.8cm}p{4.5cm}X p{4.0cm}}
\toprule
\textbf{Form of mediation} & \textbf{Role in the activity} & \textbf{Typical implementations} & \textbf{References} \\
\midrule
Producing and adapting materials & Transformed family context or learning targets into artifacts used in an activity & Stories, questions, explanations, images, curricula, and activity scenarios & \cite{chen2025characterizing,lee2024open,xu2025accompany,antony2026ella,liu2024he,kim2025bridging,rifat2026homeroom,song2026ambiguousloss} \\
Structuring dialogue and reflection & Shaped the sequence and content of participation during an activity & Conversational prompts, follow-up questions, feedback, and multi-agent perspectives & \cite{dietz2024contextq,wang2025charactercritique,he2025storypal,choi2025aacesstalk,yang2026autiverse,seo2024chacha,shi2025ineedyourhelp} \\
Interpreting context and providing guidance & Converted information about a learner, interaction, or media object into observations or recommendations & Parent guidance, content assessment, interaction analysis, journaling analysis, and multimodal interpretation & \cite{liu2025bricksmart,dangol2025want,nawshin2026well,shi2026towards,kalanadhabhatta2024playlogue,yang2026autiverse,shi2025ineedyourhelp,moon2026promises} \\
Connecting external knowledge and values & Brought professional, institutional, curricular, cultural, or household knowledge into family activity & Advocacy support, health communication, curriculum alignment, information access, and value-sensitive generation & \cite{zaidi2025sociotechnical,seo2025enhancing,rifat2026homeroom,kaur2025familyplanning,wu2026warmsteward,chheda2025artinsight,petsolari2024socio,wen2025families,zanardi2026dinner} \\
\bottomrule
\end{tabularx}
\end{table}

\textbf{Producing and adapting materials turned family context into resources for activity.} Story-reading systems personalized generated material around children's interests~\cite{chen2025characterizing} or vocabulary profiles~\cite{lee2024open}. Other systems incorporated family experiences into stories across geographic separation~\cite{xu2025accompany} or generated dialogue around parent-selected vocabulary~\cite{antony2026ella}. Family members used generated material in multimaterial storymaking~\cite{liu2024he}, intergenerational co-creation~\cite{kim2025bridging}, and homeschooling curriculum development~\cite{rifat2026homeroom}. Culture Link generated composite images that prompted families' own stories about home~\cite{song2026ambiguousloss}. That account supports a role for generated images in interaction; it does not by itself identify an LLM contribution. Generation mediated the educational object by creating a provisional resource whose relevance was completed through family participation.

\textbf{Structuring dialogue and reflection organized the sequence of participation.} ContextQ generated questions for parent--child co-reading~\cite{dietz2024contextq}, while CharacterCritique introduced contrasting perspectives for children to evaluate~\cite{wang2025charactercritique}. StoryPal prompted explanation and prediction during dialogic reading~\cite{he2025storypal}. Other systems supported communication with minimally verbal children~\cite{choi2025aacesstalk}, elicited adolescents' everyday narratives~\cite{yang2026autiverse}, guided emotional reflection~\cite{seo2024chacha}, or sustained discussion through interpreted sandplay artifacts~\cite{shi2025ineedyourhelp}. Generated questions and responses shaped participation, while children supplied experiences and interpretations and adults extended the exchange through relational or professional judgment.

\textbf{Interpreting context and providing guidance made selected features of family activity available for decision-making.} BrickSmart recommended spatial language for family block play~\cite{liu2025bricksmart}. Home-practice research elicited parent needs and expert feedback on proposed AI guidance~\cite{dangol2025want}. Other systems assessed video appropriateness~\cite{nawshin2026well} or analyzed parent--child joint attention~\cite{shi2026towards}. Multimodal implementations represented adult--child conversation~\cite{kalanadhabhatta2024playlogue}, adolescent journals~\cite{yang2026autiverse}, participant-created sandplay scenes~\cite{shi2025ineedyourhelp}, and family case records~\cite{moon2026promises}. These implementations served different users. Family-facing tools supported interpretation by families or professionals, whereas conversation and case-record analyses also served researchers; their evaluation does not establish that families acted on the outputs.

\textbf{Connecting external knowledge and values brought resources from beyond the household into family activity.} Special-education advocacy systems translated knowledge about IEP processes for parents~\cite{zaidi2025sociotechnical}. ARCH proposed a role in pediatric communication and was appraised by care experts~\cite{seo2025enhancing}; Homeroom related family values to curriculum standards~\cite{rifat2026homeroom}. Other systems supported access to family-planning knowledge~\cite{kaur2025familyplanning} or intergenerational learning about AI~\cite{wu2026warmsteward}. Families supplied accessibility expectations~\cite{chheda2025artinsight}, parenting values~\cite{petsolari2024socio}, household boundaries~\cite{wen2025families}, and perspectives on bias~\cite{zanardi2026dinner}. A single technical capability could assume several educational roles. Content generation could support child learning, provide parents with teaching materials, or create a shared object for family discussion. Analysis could support reflection or expand evaluative authority.

\paragraph{Research gaps for RQ2.}
\textit{RG2a: Personalization is studied as immediate adaptation more often than development over time.} Systems personalized stories, vocabulary, prompts, or curricula using interests, performance, parent input, or family values~\cite{chen2025characterizing,lee2024open,rifat2026homeroom,antony2026ella}. These implementations adapted support to information available at a particular point in the interaction. The studies provide less evidence about how learner representations change with development, how mistaken profiles are repaired, or how family members negotiate the information used for personalization. Research on personalized education shows that useful adaptation depends on updating learner representations as characteristics and performance change~\cite{tetzlaff2021developing}. Research on developmentally appropriate parental controls further demonstrates that one design cannot accommodate changing children, families, and contexts~\cite{dumaru2025one}. Longitudinal evidence could establish how children, family members, and systems share responsibility for updating learner representations. It could also explain how families resolve disagreements about those representations as educational needs and responsibilities change.

\textit{RG2b: The use of LLM-generated analyses in educational decisions remains underexamined.} Some studies used LLMs to transcribe interaction, analyze parent--child behavior, assess whether content was appropriate for children, or interpret the development of a family case~\cite{kalanadhabhatta2024playlogue,shi2026towards,nawshin2026well,moon2026promises}. Their evaluations established analytic or benchmark performance. When an output is intended to support family education, technical performance alone does not show whether a parent or professional can understand the analysis and use it to inform a decision. Human-centered research on learning analytics shows that an analytic result acquires educational significance when participants can interpret it and use it in decision-making~\cite{alfredo2024human}. Evidence is still needed on how participants interpret an analysis and decide whether to accept, revise, or reject it. Following that decision into practice would show whether the output changes subsequent educational guidance.

\subsection{Findings 3: How Participation and LLM Use Are Organized}
\subsubsection{LLM Roles and Human Responsibilities (RQ3.1)}
\label{Findings-RQ3.1}

\begin{figure*}[!b]
    \centering
    \includegraphics[width=0.98\textwidth]{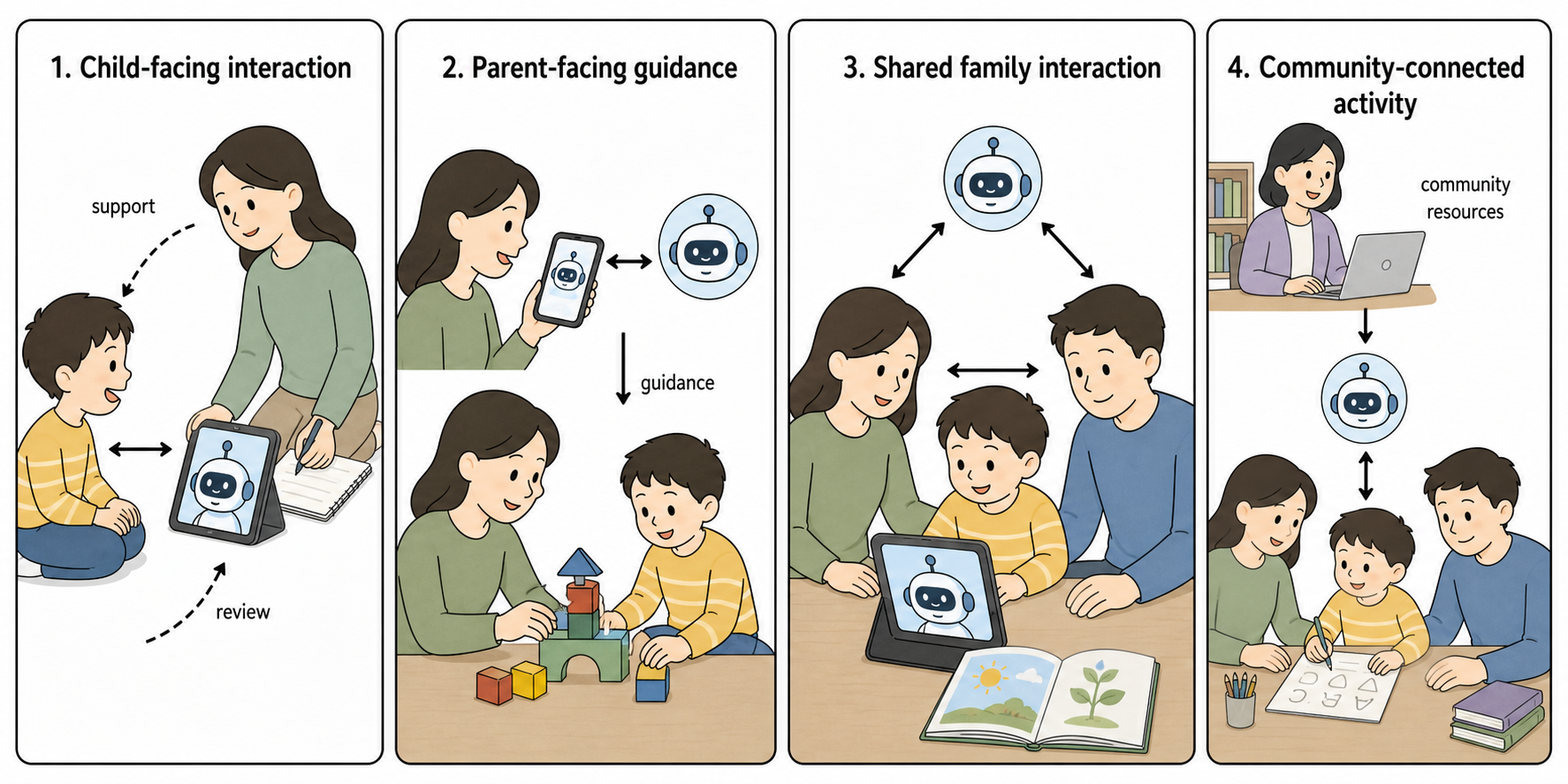}
    \caption{Recurring arrangements of LLM mediation in family education. The system may interact directly with a learner, support a family member who guides learning, participate in shared family activity, or connect family subjects with external communities.}
    \Description{Four panels compare who receives or uses an LLM contribution: a child interacts with a system; a parent receives guidance while supporting a child; family members jointly use generated material; and a family activity connects with an external professional. The panels illustrate interaction arrangements, not measured outcomes or a progression between stages.}
    \label{fig:llm-interaction-forms}
\end{figure*}

Table~\ref{tab:division-of-labor-arrangements} summarizes reported and proposed task allocations among LLM-based systems, family members, and external communities. These allocations identify responsibilities; they do not establish a net change in workload. Figure~\ref{fig:llm-interaction-forms} visualizes how the four arrangements position the system in relation to learners, family members, and external communities.

\begin{table}[H]
\centering
\small
\caption{Roles described or proposed in family-facing designs. The examples include implemented systems and design requirements; task allocations do not all represent observed use.}
\label{tab:division-of-labor-arrangements}
\begin{tabularx}{\textwidth}{p{2.8cm}p{3.6cm}X p{4.0cm}}
\toprule
\textbf{Arrangement} & \textbf{System contribution described} & \textbf{Human responsibilities described} & \textbf{References} \\
\midrule
\textbf{Direct interaction with a learner} & Generated prompts, stories, feedback, or conversational responses & Learners disclosed, responded, interpreted, corrected, or created; adults set goals and supervised use & \cite{chen2025characterizing,he2025storypal,yang2026autiverse,seo2024chacha,antony2026ella} \\
\textbf{Support for a family member who guided learning} & Produced recommendations, draft materials, explanations, or preliminary assessments & Family members evaluated fit, adapted outputs, guided learning, and made educational decisions & \cite{choi2025aacesstalk,liu2025bricksmart,dangol2025want,nawshin2026well,rifat2026homeroom,zaidi2025sociotechnical} \\
\textbf{Participation in shared family activity} & Supplied prompts or provisional artifacts for joint use & Family members contributed situated knowledge, negotiated meaning, and maintained relational participation & \cite{zhao2025youthcare,long2022family,chheda2025artinsight,liu2024he,kim2025bridging,dietz2024contextq,song2026ambiguousloss,wu2026warmsteward} \\
\textbf{Connection with external communities} & Translated, summarized, or represented professional and institutional knowledge & Families supplied context; professionals and institutions exercised domain authority and accountability & \cite{dangol2025want,seo2025enhancing,shi2025ineedyourhelp} \\
\bottomrule
\end{tabularx}
\end{table}

Across the four arrangements, studies described different allocations of educational tasks. \textbf{Direct interaction with a learner.} Systems supplied reading material~\cite{chen2025characterizing}, dialogic prompts~\cite{he2025storypal}, or support for emotional reflection~\cite{seo2024chacha}. Children contributed experiences and interpretations, with adult guidance arranged around the interaction.

\textbf{Support for a family member who guided learning.} Communication scaffolds~\cite{choi2025aacesstalk} and spatial-language guidance~\cite{liu2025bricksmart} supplied resources for parents to interpret and adapt. Home speech-practice research instead examined desired AI support through parent interviews and expert review of concepts~\cite{dangol2025want}.

\textbf{Participation in shared family activity.} During media evaluation~\cite{zhao2025youthcare}, AI-literacy activities~\cite{long2022family}, and intergenerational co-creation~\cite{kim2025bridging}, family members connected generated material with their experiences and negotiated its use.

\textbf{Connection with external communities.} The proposed roles in home speech practice~\cite{dangol2025want} and the expert-appraised ARCH prototype~\cite{seo2025enhancing} connected family concerns with professional expertise. Sandplay interpretation provided another professional-facing contribution~\cite{shi2025ineedyourhelp}. Retrospective child-welfare analysis~\cite{moon2026promises} concerned institutional information work. These cases require separate examination of intended responsibilities and what people actually did with system outputs.

\paragraph{Research gaps for RQ3.1.}
\textit{RG3.1a: Verification and repair labor remains weakly documented.} Parents were expected to review generated content, adapt recommendations, or retain final decisions in reading, home practice, media assessment, and homeschooling~\cite{ho2025set,dangol2025want,nawshin2026well,rifat2026homeroom}. Studies reported less consistently who detected inappropriate output, repaired mistaken interpretations, or explained system failures to a child. Research on collaboration between parents in family education shows that families distribute and negotiate educational work~\cite{wang2026division}. Documenting this redistribution requires evidence about the work introduced by system use. Such evidence can show which family members check outputs, correct errors, coordinate participation, and manage the interaction's emotional consequences.

\textit{RG3.1b: Redistribution of labor is examined mainly through parent--child configurations.} Parents supplied communication context~\cite{choi2025aacesstalk}, scaffolded family learning~\cite{liu2025bricksmart}, monitored children's AI use~\cite{driscoll2026understanding}, and connected family activity with educational institutions~\cite{zaidi2025sociotechnical}. Evidence is much thinner for siblings, several caregivers, extended families, and changing participation across households. Sibling teaching demonstrates that children can assume educational work within family activity~\cite{maynard2002cultural}. Collaboration between parents further shows that adults negotiate how educational labor is divided and coordinated within the family~\cite{wang2026division}. Comparative studies should examine whether LLM use concentrates work in one caregiver or enables responsibility to move among family members and households.

\textit{RG3.1c: Responsibility across family and professional boundaries remains unresolved.} Community-connected systems brought clinical and educational knowledge into family activity through pediatric communication~\cite{seo2025enhancing} and therapeutic interpretation~\cite{shi2025ineedyourhelp}. Analysis of child-welfare case records raised related questions about practitioner judgment and institutional accountability~\cite{moon2026promises}, but did not establish how families used the outputs. Across these settings, professional review and family contextualization must be examined as distinct tasks. Boundary-crossing research shows that moving knowledge between communities requires interpretation across differences in practice and authority~\cite{akkerman2011boundary}. Research on special-education partnerships further identifies fragmented roles and uneven collaboration as consequential problems across institutional boundaries~\cite{ko2021systematic}. Responsibility for LLM-supported guidance concerns both educational coordination and technical accuracy. Future research should trace who interprets, approves, contests, and remains accountable for guidance as it moves between households and institutions.

\subsubsection{Reported Rules for Authority, Acceptability, and Information Boundaries (RQ3.2)}
\label{Findings-RQ3.2}

Table~\ref{tab:rule-domains} summarizes reported controls and expectations concerning participation, the acceptability of LLM contributions, and information boundaries in family education activities. Some describe implemented controls; others describe desired boundaries elicited in research.

\begin{table}[H]
\centering
\small
\caption{Rule domains governing participation and LLM use in family educational activity.}
\label{tab:rule-domains}
\begin{tabularx}{\textwidth}{p{3.0cm}p{4.5cm}X p{4.0cm}}
\toprule
\textbf{Rule domain} & \textbf{What the rules governed} & \textbf{Recurring concerns} & \textbf{References} \\
\midrule
Participation and authority & Who could use the system, intervene, review outputs, and make final decisions & Child autonomy, adult supervision, readiness, system initiative, and professional authority & \cite{ho2025set,driscoll2026understanding,xie2026understanding,chen2025characterizing,nawshin2026well,rifat2026homeroom,yang2026autiverse,seo2024chacha,yoo2026creativecollaboration} \\
Acceptability of system contributions & Conditions under which generated content or guidance could enter the activity & Developmental fit, reliability, emotional safety, authorship, cultural fit, and family values & \cite{sun2024exploring,wang2025charactercritique,he2025storypal,antony2026ella,dangol2025want,seo2025enhancing,wester2024facing,shi2025ineedyourhelp,xu2025accompany,petsolari2024socio,song2026ambiguousloss,zanardi2026dinner} \\
Information boundaries and accountability & What information could be collected, disclosed, shared, or acted upon across settings & Privacy, parental visibility, child disclosure, confidentiality, documentation, and institutional responsibility & \cite{yang2026autiverse,driscoll2026understanding,seo2024chacha,zaidi2025sociotechnical,seo2025enhancing,moon2026promises} \\
\bottomrule
\end{tabularx}
\end{table}

\textbf{Participation rules coordinated child autonomy and adult authority.} SET-PAiREd allowed parents to adjust their involvement in children's learning with an educational robot~\cite{ho2025set}. Parents also specified oversight preferences for children's chatbot interactions~\cite{driscoll2026understanding} and described support appropriate to children's readiness for independent AI use~\cite{xie2026understanding}. Adults reviewed personalized stories~\cite{chen2025characterizing}, assessed video appropriateness~\cite{nawshin2026well}, and edited homeschooling materials~\cite{rifat2026homeroom}. Child-facing systems supported choices about narrative disclosure~\cite{yang2026autiverse}, emotional expression~\cite{seo2024chacha}, and creative contribution~\cite{yoo2026creativecollaboration}. These rules distributed authority across child agency, adult judgment, and system initiative.

\textbf{Acceptability rules connected system participation with situated human judgment.} Families evaluated whether storytelling support matched children's needs~\cite{sun2024exploring}, whether generated perspectives supported critical interpretation~\cite{wang2025charactercritique}, and whether prompts fitted dialogic reading~\cite{he2025storypal}. Vocabulary-learning robots also required content suited to the selected learning target~\cite{antony2026ella}. Professional judgment shaped home speech practice~\cite{dangol2025want}, pediatric communication~\cite{seo2025enhancing}, family health-information mediation~\cite{wester2024facing}, and therapeutic interpretation~\cite{shi2025ineedyourhelp}. Cultural and relational expectations shaped stories for geographically separated families~\cite{xu2025accompany}, imagined parenting support~\cite{petsolari2024socio}, narratives of displacement~\cite{song2026ambiguousloss}, and family discussion of bias~\cite{zanardi2026dinner}. Across evaluations and elicited expectations, developmental fit, reliability, emotional safety, and value alignment were recurring concerns.

\textbf{Information rules connected household privacy with external accountability.} AI-guided journaling created choices about adolescent disclosure~\cite{yang2026autiverse}. Parents considered access to children's chatbot interactions~\cite{driscoll2026understanding}, while emotional-reflection systems made children's disclosures relevant to parental support~\cite{seo2024chacha}. Community-connected systems introduced documentation requirements in special education~\cite{zaidi2025sociotechnical}, and confidentiality in pediatric communication~\cite{seo2025enhancing}, while retrospective analysis of child-welfare records raised institutional accountability concerns~\cite{moon2026promises}. Household preferences, platform practices, and external obligations jointly shaped information boundaries.

Rules specified the relationship an LLM-based system could assume and distributed authority among children, adults, professionals, and institutions. Many studies elicited desired rules through interviews, workshops, or design probes. These results characterize expected boundaries and provide a basis for studying how families establish and revise rules through sustained use.

\paragraph{Research gaps for RQ3.2.}
\textit{RG3.2a. Anticipated rules are documented more often than rules enacted during sustained use.} Studies of children's storytelling systems elicited parents' expectations for supervision and acceptable system roles~\cite{sun2024exploring}. Design fictions surfaced boundaries around AI parenting support~\cite{petsolari2024socio}. Interviews identified parents' preferred forms of access and intervention for children's chatbot use~\cite{driscoll2026understanding}, while household probes elicited expectations about the authority of generative safety agents~\cite{wen2025families}. These studies established rules that participants wanted but provided little evidence of how families applied, contested, or revised them during everyday use. Research on developmentally appropriate parental controls shows that governance requirements change with children's development and the context of use~\cite{dumaru2025one}. Studies conducted across repeated use can examine how family relationships, educational purposes, and children's growing experience alter the rules that initially appeared acceptable.

\textit{RG3.2b. Child privacy and agency remain in tension with parental oversight.} AI-guided journaling allowed autistic adolescents to record experiences that caregivers and professionals might later interpret~\cite{yang2026autiverse}. Emotional reflection systems similarly prepared children's disclosures for possible discussion with parents~\cite{seo2024chacha}. Self-directed learning research positioned parental involvement as adjustable according to children's readiness~\cite{xie2026understanding}, while chatbot studies elicited parents' preferences for viewing conversations and intervening in risky interactions~\cite{driscoll2026understanding}. The corpus identifies both children's interest in controlling expression and adults' responsibility for safety, but it offers limited evidence about how families resolve disagreements over visibility. Research on joint media engagement treats children and parents as active participants whose interaction shapes the meaning and consequences of media use~\cite{yu2024joint}. Examining access, disclosure, review, and intervention as negotiated rules can show how privacy and oversight are balanced when educational data are sensitive.

\textit{RG3.2c. Conflicts between household rules and external obligations are seldom examined.} Special education systems introduced procedural requirements and professional authority into parents' advocacy work~\cite{zaidi2025sociotechnical}. Pediatric communication systems brought clinical confidentiality and responsibility into exchanges involving children and parents~\cite{seo2025enhancing}. Public service applications placed family information within institutional documentation and accountability practices~\cite{moon2026promises}. The corpus describes these obligations separately and provides little evidence about what happens when they conflict with household values or platform rules. Boundary-crossing research shows that differences in practice, responsibility, and authority complicate collaboration across communities~\cite{akkerman2011boundary}. Reviews of special education partnerships further show that institutional boundaries shape communication, coordination, and stakeholders' capacity to address shared educational needs~\cite{ko2021systematic}. Studying LLM use where household and institutional rules meet can reveal whose interpretation governs disputed action, how families and professionals contest a decision, and who remains accountable for its consequences.


\section{Discussion}
\label{Discussion}

Across the reviewed designs, generation supplies a resource whose educational role depends on who uses it and who may revise or reject it. A co-reading question enters a parent--child exchange~\cite{dietz2024contextq}; a homeschooling draft is judged against curricular expectations and family values~\cite{rifat2026homeroom}; and advice about home speech practice must fit a family's routines and professional guidance~\cite{dangol2025want}. Activity theory makes these relationships comparable by connecting the educational object with mediation, division of labor, and rules. The resulting design questions concern who can shape the activity, who carries the checking work, and how decision authority changes. The following agenda develops those questions from the synthesis.

\subsection{Rethinking the Boundaries of LLM-Based Family Education}

The corpus concentrates on child--parent relationships: 44 of 53 studies were coded around that relationship, including parent-only investigations about a child; one centered grandparents and grandchildren, and none centered siblings. Educational objects clustered around language, narrative, AI literacy, and emotional or relational development. Logical reasoning~\cite{liu2025bricksmart} and self-directed learning~\cite{xie2026understanding} received less attention. These are gaps in the retrieved literature; they do not establish whether LLMs have broadened family education relative to earlier technologies.

Family-learning research gives reasons to examine these gaps. Siblings, grandparents, and other relatives contribute different forms of support~\cite{muminova2025family,marjoribanks2005family,fuoco2024parent,maynard2002cultural,davies2019sticky,lee2024home}. Home literacy and numeracy~\cite{melhuish2008effects,skwarchuk2014formal,soto2020identifying}, science inquiry~\cite{national2009learning}, and homework routines~\cite{hill2009parental} also involve resources and activities beyond dialogue. A focus on generated conversation can overlook learning through physical exploration and changing responsibilities.

Future designs could preserve who contributed a question, correction, strategy, or family memory, then use this attribution to coordinate participation. Intergenerational storymaking~\cite{kim2025bridging} and reciprocal GenAI assistance between younger and older relatives~\cite{wu2026warmsteward} provide starting points. Mathematical and scientific activities could connect prompts with manipulatives, drawings, or measurements and invite participants to compare reasoning before receiving an explanation, as spatial guidance in BrickSmart illustrates~\cite{liu2025bricksmart}. Homework analysis~\cite{gao2025homework} could inform ways to make changes in assistance visible, while self-directed learning designs could support explicit handoffs of responsibility~\cite{xie2026understanding}.

Several users sharing an interface does not establish shared participation. Expanding access can expose a child's learning history or private disclosures to more relatives, as chatbot moderation and journaling studies make salient~\cite{driscoll2026understanding,yang2026autiverse}. Contribution histories need selective visibility and revisable permissions~\cite{dumaru2025one}; Section~\ref{sec:family-governance} develops this issue. Repeated-use studies should examine whose contributions shape the activity and who checks or repairs outputs. Interests, self-regulation, and parental assistance change over time~\cite{zimmerman2002becoming,silinskas2020responsive,waters2026transactional}. Bedtime storytelling~\cite{xu2025accompany} and homeschooling~\cite{rifat2026homeroom} offer settings to follow those changes, including correction, rejection, and abandonment of generated material.

\subsection{Evaluating Long-Term Personalization Across Changing Family Educational Needs}

The reviewed systems personalize support using current information: StoryMate draws on children's interests~\cite{chen2025characterizing}, ELLA on selected vocabulary~\cite{antony2026ella}, and Homeroom on parent input and family values~\cite{rifat2026homeroom}. These implementations provide less evidence about whether adaptation remains appropriate as learners develop and family purposes change. Adaptive-education research likewise identifies weak empirical justification for how learner models develop~\cite{wang2025learnermodel}.

Open learner model research proposes inspection, challenge, and additional evidence as ways to repair inaccurate or outdated representations~\cite{bull2016negotiated}. Family education adds the possibility that a child and caregiver interpret the same need differently. Systems could show the evidence behind personalized outputs, communicate uncertainty, and let participants propose corrections without silently overriding one another. Longitudinal evaluation should examine when profiles become outdated, how disagreements are resolved, and whether revisions improve alignment with changing educational goals. The issue is who can maintain a useful representation of the learner over time, as well as what content the system can generate from it.

\subsection{Rethinking Assistance as a Redistribution of Family Labor}

The reviewed designs assign generation and preliminary analysis to systems while retaining human judgment. This suggests a redistribution of tasks, but does not establish a net reduction in family workload. Research on GenAI collaboration describes the continuing effort to sustain assistance as articulation work~\cite{boulusrodje2024hidden}. The family cases locate similar responsibilities: parents interpret BrickSmart's guidance during play~\cite{liu2025bricksmart}, identify how speech-practice support could fit routines~\cite{dangol2025want}, and review Homeroom materials against curricular expectations and family values~\cite{rifat2026homeroom}.

CharacterCritique provides a useful contrast: agent-supported reading involved fewer parent--child dialogue rounds but more discussion topics than traditional reading~\cite{wang2025charactercritique}. Conversational volume alone cannot indicate the educational value of parental participation. The relevant question is what guidance parents contribute and when they intervene.

Generation may save preparation while requiring additional checking, correction, and coordination. Evaluations should measure these activities alongside usability and learning outcomes: time spent reviewing, the substance of corrections, handoffs before use, and the responsibility of accepting or rejecting guidance. This would test the balance of effort saved and introduced, which the current evidence does not establish.

Making work visible can also concentrate it on the person already carrying most family responsibility. Homecare research shows how technology can expose coordination demands without supporting their negotiation~\cite{renyi2022complexity}. Family education systems could let participants assign, decline, defer, and hand off review tasks. They should distinguish system output, learner response, caregiver revision, and professional approval, keeping unresolved recommendations visibly unresolved. Evaluation should trace who receives requests and performs repairs.

When guidance crosses household and professional boundaries, provenance and authority also matter. Pediatric communication~\cite{seo2025enhancing} and special-education advocacy~\cite{zaidi2025sociotechnical} connect family knowledge with clinical or institutional expertise. Designs should preserve sources and distinguish generated interpretations from authoritative decisions. Explicit review states could show who has considered a recommendation, what remains disputed, and who can authorize action. Such mechanisms would support task negotiation while keeping responsibility visible.

\subsection{From Configurable Controls to Negotiated Family Governance}

\label{sec:family-governance}
The corpus documents expectations about access, supervision, acceptable content, professional authority, and sharing. Many came from interviews, workshops, or design probes; they describe anticipated rules more than rules enacted and revised during sustained use. A settings panel can express an initial preference, but cannot settle conflicts among a child's privacy, a caregiver's responsibility, platform moderation, and institutional requirements.

Readiness for chatbot interaction~\cite{driscoll2026understanding} and self-directed learning~\cite{xie2026understanding} changes what oversight families consider appropriate. Parental-control research similarly finds variation across children, activities, and developmental stages~\cite{dumaru2025one}. Systems could represent rules as revisable agreements: who proposed them, which activity or data they cover, and when they should be reconsidered. Evaluation should track whose objections alter a rule and whether growing experience brings greater decision authority.

Selective disclosure is one concrete test. Journaling~\cite{yang2026autiverse} and emotional-reflection systems~\cite{seo2024chacha} connect children's accounts with possible adult interpretation. Building on joint media engagement~\cite{yu2024joint}, children could preview what will be shared, caregivers could explain requests for access, and families could contest inferences or correct records. Designs should distinguish routine monitoring from safety escalation and document unresolved disagreement. Neither unrestricted access nor comprehensive parental visibility should be assumed appropriate for every activity.

Household decisions can also meet professional rules in special-education advocacy~\cite{zaidi2025sociotechnical} and pediatric communication~\cite{seo2025enhancing}. Boundary-crossing scholarship helps explain the different standards and responsibilities involved~\cite{akkerman2011boundary}. External requirements need attributable sources and scope; generated interpretations and family adaptations need to remain distinguishable from professional decisions. The review states proposed above could make these handoffs and disagreements inspectable.

Finally, families need intelligible reasons for media judgments~\cite{nawshin2026well} and may share educational resources within trusted communities~\cite{rifat2026homeroom}. Selective sharing, editable value statements, and visible moderation rationales are candidate mechanisms to investigate. Their value should be assessed through families' ability to understand, challenge, and repair decisions over time, including whose voice is heard when preferences conflict.

\subsection{Limitations of the Review}

The venue-based search bounds this review to the selected HCI publication outlets. It may omit relevant work in education, communication, clinical research, and other disciplines, as well as publications outside the selected venues. Metadata searches can also miss studies whose titles, abstracts, and keywords do not use the selected technology, education, or family terms. The reported distributions characterize the retrieved corpus, rather than the prevalence of LLM use in families or the full family-education literature.

The corpus combines evaluated systems, investigations of needs and expectations, and retrospective computational analyses. These studies support different claims: a desired feature is not an observed benefit, and model performance does not establish a change in family learning or interaction. Our synthesis maps these contributions without estimating a common intervention effect. Its conclusions also depend on what the primary studies reported about participants, practices, and evaluation. Representing each study through a focal activity configuration facilitates comparison but can underrepresent secondary relationships and changes within an activity. Finally, the AODM lens foregrounds relations among participants, tools, purposes, and responsibilities; other analytical frameworks could expose different patterns. The proposed design directions should consequently be treated as interpretations to investigate, particularly through sustained family use.

\section{Conclusion}
This review has examined 53 HCI studies of LLM-based family education using an activity-theoretical framework operationalized through AODM. The analysis connected family participants and educational objects with the technologies, labour, and rules that organized each activity. Current research centers mainly on child--parent interaction and on language, narrative, AI literacy, and relational development, leaving other family relationships, educational domains, and sustained routines with a smaller evidence base.

Across the corpus, LLM capabilities acquired educational meaning through their implementation and situated use. Systems supplied provisional materials, structured interaction, interpreted activity, and connected households with external knowledge, while family members judged how these contributions fitted the learner and situation. LLM use consequently redistributed educational labour while leaving parents responsible for much of the supervision and correction and preserving professional authority over clinical or institutional knowledge. Participation, acceptability, and information rules further determined what systems could contribute and who could act on their outputs.

The available evidence remains limited by the concentration on child--parent dyads and short deployments. Future research should examine diverse family configurations over time and follow LLM-generated analyses into the educational decisions they inform. Doing so can move HCI beyond evaluating individual systems toward explaining how families negotiate participation, responsibility, and authority when LLMs enter educational activity.

\section*{AI Use Disclosure}
OpenAI's GPT was used to support language editing, consistency checks, and drafting revisions to evidence descriptions and review limitations. The authors are responsible for study selection, coding, interpretation, and the final manuscript. Figure~2 was produced with the assistance of OpenAI Codex based on the concepts, relationships, and intended meaning specified by the authors, who reviewed and approved the resulting visualization.

\newpage
\bibliographystyle{ACM-Reference-Format}
\bibliography{01_PQE}

\newpage
\section*{Appendix 1: Search Information and Coding Framework}
The selected venues were searched through the six publisher platforms listed in Table~\ref{tab:search-terms}. Search syntax was adapted to the functions and limits of each platform. ScienceDirect limited each field to eight Boolean connectors, so the complete search logic was executed through nine segments and the results were merged and deduplicated. Frontiers was searched with the technology terms and restricted to \textit{Frontiers in Virtual Reality}; educational purpose and family participation were assessed during screening.

\newcommand{\fullsearchstring}{(``LLM'' OR ``large language model'' OR ``generative AI'' OR ``chatbot'' OR ``conversational agent'' OR ``agent'') AND (``education'' OR ``learning'' OR ``pedagogy'' OR ``teaching'') AND (``family'' OR ``families'' OR ``parents'' OR ``home'' OR ``caregiver'' OR ``child'' OR ``children'' OR ``sibling'' OR ``grandparent'')}

{\small
\begin{xltabular}{\textwidth}{p{3.4cm}X}
\caption{Search platforms and corresponding executed keyword queries.}
\label{tab:search-terms}\\
\toprule
\textbf{Database or platform} & \textbf{Keyword} \\
\midrule
\endfirsthead
\multicolumn{2}{l}{\textit{Table~\thetable{} continued from previous page}} \\
\toprule
\textbf{Database or platform} & \textbf{Keyword} \\
\midrule
\endhead
\midrule
\multicolumn{2}{r}{\textit{Continued on next page}} \\
\endfoot
\bottomrule
\endlastfoot
ACM Digital Library; IEEE Xplore; SpringerLink; Taylor \& Francis Online & \fullsearchstring \\
ScienceDirect & Nine segments of $T_i$ AND $E$ AND $F_j$ were executed for all combinations of $i,j \in \{1,2,3\}$. The sets were: $T_1$ = (``LLM'' OR ``large language model''); $T_2$ = (``generative AI'' OR ``chatbot''); $T_3$ = (``conversational agent'' OR ``agent''); $E$ = (``education'' OR ``learning'' OR ``pedagogy'' OR ``teaching''); $F_1$ = (``family'' OR ``families'' OR ``parents''); $F_2$ = (``home'' OR ``caregiver'' OR ``child''); and $F_3$ = (``children'' OR ``sibling'' OR ``grandparent''). Each segment contained no more than eight Boolean connectors. \\
Frontiers & (``LLM'' OR ``large language model'' OR ``generative AI'' OR ``chatbot'' OR ``conversational agent'' OR ``agent''). Journal restriction: \textit{Frontiers in Virtual Reality}. \\
\end{xltabular}
}

\AODMCodebook

\newpage
\section*{Appendix 2: Paper-Level Coding}

\noindent\textit{Coding note.} NR indicates that the activity component was not explicitly reported in the primary study. Paper identifiers correspond to the corpus used throughout the review. Reported findings include learning and interaction findings, participant perceptions, design requirements, and technical results. These evidence types should not be interpreted as equivalent educational outcomes. Subjects may be participants or actors represented in records, rather than direct users of an LLM.

Tables~\ref{tab:appendix-activity-context} and~\ref{tab:appendix-activity-organization} provide complementary views of the same 53 studies. Table~\ref{tab:appendix-activity-context} reports the activity context and findings. Table~\ref{tab:appendix-activity-organization} reports technological mediation and social organization, including tools, rules, and divisions of labour. Paper identifiers are consistent across the two tables.

\begin{xltabular}{\textwidth}{@{}p{0.55cm} p{3.0cm} p{2.5cm} p{2.5cm} X p{0.65cm}@{}}
\caption{Activity objects, participants, communities, and reported findings in the primary studies.}
\label{tab:appendix-activity-context}\\
\toprule
\textbf{ID} & \textbf{Object} & \textbf{Subjects} & \textbf{Community} & \textbf{Reported findings} & \textbf{Ref.} \\
\midrule
\endfirsthead
\multicolumn{6}{@{}l}{\textit{Table~\thetable{} continued from previous page}} \\
\toprule
\textbf{ID} & \textbf{Object} & \textbf{Subjects} & \textbf{Community} & \textbf{Reported findings} & \textbf{Ref.} \\
\midrule
\endhead
\midrule
\multicolumn{6}{r@{}}{\textit{Continued on next page}} \\
\endfoot
\bottomrule
\endlastfoot
P1 & Personalized story reading and interaction & Children, parents, education experts & Family and education experts & Interaction findings: personalization and engagement. Design findings: requirements for story reading & \cite{chen2025characterizing} \\
P2 & Everyday reciprocal communication & Minimally verbal autistic children and parents & Family and autism-support context & More frequent conversation, smoother turn-taking, child agency, and parental efficacy & \cite{choi2025aacesstalk} \\
P3 & Spatial-language learning in family block play & Children and parents & Family learning context & Spatial-language use, learning, engagement, and parental guidance & \cite{liu2025bricksmart} \\
P4 & Joint evaluation of children's video content & Children and parents & Household media context & Shared preferences, content judgments, consensus, and joint media engagement & \cite{zhao2025youthcare} \\
P5 & Social-emotional reflection during media use & Children and parents & Family media context & Emotional reflection and parent-child interaction & \cite{shen2025easel} \\
P6 & Child-appropriate storytelling and reading support & Preschool children and parents & Family reading context & Design requirements: appropriate, engaging, and parent-supported AI interaction & \cite{sun2024exploring} \\
P7 & Parent-child homework interaction & Children and parents & Household homework context & Analytic findings: emotions, behaviours, conflicts, and support opportunities in homework records & \cite{gao2025homework} \\
P8 & Children's problem solving with conversational agents & Children and knowledgeable others & Family or learning support context & Cognitive scaffolding, problem solving, and engagement & \cite{figueiredo2025designing} \\
P9 & Analysis of adult-child play conversations & Children and adults & Family play and research community & Technical evaluation: diarization, transcription, and conversational coding benchmarks & \cite{kalanadhabhatta2024playlogue} \\
P10 & Parent involvement in robot-assisted learning & Children, parents, and robot & Family learning context & Appropriate content, parental involvement, and coordinated learning roles & \cite{ho2025set} \\
P11 & Critical thinking during family story reading & Children and parents & Family reading context & Critical interpretation, engagement, and family critical-thinking education & \cite{wang2025charactercritique} \\
P12 & Child-robot collaborative storytelling & Children, adults, and robot & Family or supported storytelling context & Engagement, recall, participation, and vocabulary learning & \cite{olutunbi2026leads} \\
P13 & Family learning about AI literacy & Children and adult family members & Family and informal learning community & Family dialogue and understanding of AI-literacy competencies & \cite{long2022family} \\
P14 & Home speech-therapy practice & Children and parents & Family and speech-language pathology community & Parent needs and expert feedback on proposed support; no deployed-system learning outcome & \cite{dangol2025want} \\
P15 & Artwork access and family interpretation & Mixed visual-ability family members & Family and cultural-art context & Accessible artwork engagement and family conversation & \cite{chheda2025artinsight} \\
P16 & Generative AI in elementary literacy education & Students, parents, and teachers & Families and schools & Adaptable materials, ideation, feedback, agency, and responsible adoption & \cite{han2024teachers} \\
P17 & Family expressive-arts therapy through storymaking & Children, parents, and therapists & Family and therapeutic community & Emotional expression, shared storymaking, and therapeutic interaction & \cite{liu2024he} \\
P18 & Personalized vocabulary assessment and intervention & Children and supporting adults & Family language environment & Vocabulary profiling, assessment, and personalized intervention & \cite{lee2024open} \\
P19 & Intergenerational co-creation & Grandparents and grandchildren & Intergenerational family context & Communication, mutual appreciation, balanced participation, and connection & \cite{kim2025bridging} \\
P20 & Parent advocacy in special education & Parents and designers & Families, schools, and IEP stakeholders & Co-design findings: advocacy barriers and proposed support for parent participation in IEP processes & \cite{zaidi2025sociotechnical} \\
P21 & Parents' perceptions of adolescents' GenAI use & Parents of adolescents & Family technology and adolescent-learning context & Parent reports: concerns, perceived opportunities, and mediation requirements & \cite{eira2025parents} \\
P22 & Autistic adolescents' daily narrative expression & Autistic adolescents, caregivers, and professionals & Family and professional support community & Externalized experiences and improved caregiver-professional understanding & \cite{yang2026autiverse} \\
P23 & Meaningful dialogue during parent-child co-reading & Children and parents & Family reading context & Dialogic conversation and parental appropriation of generated questions & \cite{dietz2024contextq} \\
P24 & Generative visual storytelling for young learners & Parents, teachers, AI researchers, and young learners as intended users & Family and education design context & Design requirements: creativity, literacy, safety, and adult mediation & \cite{han2023design} \\
P25 & Dialogic reading with LLMs & Young children and parents & Family reading context & Question quality, verbal engagement, reading participation, and perceived companionship & \cite{he2025storypal} \\
P26 & Family anti-bullying learning & Children and parents & Family learning context & Anti-bullying understanding, reflection, and family participation & \cite{liu2026dollama} \\
P27 & Expert-aligned analysis of parent-child interaction & Parents, children, and speech-language pathologists & Family and clinical-research community & Technical evaluation: alignment with expert observation and judgment of joint attention & \cite{shi2026towards} \\
P28 & Parental wellbeing support & Parents and system designers & Parenting-support community & Improved interaction, user control, usability, and wellbeing-support experience & \cite{viswanathan2025interaction} \\
P29 & Bedtime storytelling in left-behind-child families & Children and remote parents & Translocal family context & Children's understanding and parent-child emotional communication & \cite{xu2025accompany} \\
P30 & Pediatric communication & Children, parents, and clinicians & Family and healthcare community & Family design requirements and expert appraisal of ARCH; no demonstrated communication improvement & \cite{seo2025enhancing} \\
P31 & Parental moderation of children's chatbot use & Children and parents & Household technology context & Elicited preferences: concern triggers, interventions, and data-access expectations & \cite{driscoll2026understanding} \\
P32 & Responsible AI in children's self-directed learning & Children and parents & Family learning context & Parent perspectives: children's readiness, autonomy, responsible use, and parental mediation & \cite{xie2026understanding} \\
P33 & Parental assessment of video appropriateness & Parents and children & Household media context & Accurate, transparent, and useful appropriateness judgments & \cite{nawshin2026well} \\
P34 & Adolescent AI literacy and use & Adolescents and parents & Family technology context & Perception measures: AI trust, perceived creepiness, literacy, and intention to use & \cite{zhang2025parental} \\
P35 & Situated AI parenting support & Parents and designers & Family and parenting-support community & Design-fiction findings: imagined benefits, frictions, and support requirements & \cite{petsolari2024socio} \\
P36 & AI acceptance for autistic children's support & Parents & Family and autism-support context & Parent reports: acceptance and behavioural intention & \cite{zhang2026parents} \\
P37 & Postnatal child-development monitoring & Parents and caregivers & Family and child-health community & Tracking, reminders, recommendations, and collaborative caregiving opportunities & \cite{talai2025towards} \\
P38 & Family health-information communication & Parents and young adults & Family health context & Disclosure, information sharing, and perceptions of robot communication styles & \cite{wester2024facing} \\
P39 & Value-aligned homeschooling & Parents and children & Family, school standards, and trusted homeschooling circles & Parental agency, aligned materials, routine support, and trusted resource sharing & \cite{rifat2026homeroom} \\
P40 & Emotional reflection and communication & Children, with parents as intended support recipients & Child emotional learning and family communication context & Emotion identification, self-expression, reflection, and preparation for sharing with parents & \cite{seo2024chacha} \\
P41 & Creative collaboration with generative AI & Children and parents & Peer collaboration and family co-design context & Social bonding, shared ownership, agency tensions, and parent-informed design requirements & \cite{yoo2026creativecollaboration} \\
P42 & Psychological communication through sandplay & Left-behind children, remote parents, therapists, and teachers & Family and mental-health support community & Psychological-state assessment, parental understanding, communication guidance, and improved parent-child interaction & \cite{shi2025ineedyourhelp} \\
P43 & Collaborative storytelling and sense of home & Children and adults in refugee families & Refugee family and resettlement community & Memory preservation, cultural transmission, family engagement, and intergenerational negotiation of belonging & \cite{song2026ambiguousloss} \\
P44 & Critical engagement with AI-generated youth culture & Adolescents, parents, and teachers & Families, peers, schools, and social-media communities & Critical awareness, responsible engagement, and intergenerational communication about AI-generated content & \cite{schiavo2026brainrot} \\
P45 & Interpretation of child-welfare case progress & Child-welfare workers and families represented in case records & Public child-welfare agency & Retrospective case analysis: detection of progress, deviations, and substantive concerns; no demonstrated family learning outcome & \cite{moon2026promises} \\
P46 & Family planning for household safety & Family members & Household safety and security context & Design expectations: threat responses, boundaries, and requirements for generative agents & \cite{wen2025families} \\
P47 & Access to family-planning information and support & Postpartum women and family members & Families and community-health context & Information access, reproductive-health understanding, agency, privacy concerns, and support needs & \cite{kaur2025familyplanning} \\
P48 & Children's mental-resilience learning & Children and supporting adults & Child-development and family-support context & Resilience assessment, reflective dialogue, and design requirements for sustained support & \cite{hu2024grow} \\
P49 & Everyday intergenerational GenAI learning & Younger and older family members & Chinese family and household context & Reciprocal learning, digital backfeeding, stewardship, confidence, and responsible GenAI use & \cite{wu2026warmsteward} \\
P50 & Learning to challenge gender stereotypes & Children and parents & Family values and critical-literacy context & Critical reflection, stereotype awareness, and parent-informed educational scenarios & \cite{hassan2026anyone} \\
P51 & Family reflection on GenAI gender bias & Children and parents & Family and informal-learning context & Awareness, curiosity, discernment, respect, pluralism, and design requirements for reflection & \cite{zanardi2026dinner} \\
P52 & Parental conditions for children's GenAI learning & Parents of K--12 children & Family and educational-access context & Parent reports: differences in perceptions, adoption, and capacity to mediate children's GenAI use & \cite{baba2026digitalinequality} \\
P53 & Early language development at home & Preschool children, parents, educators, and social robot & Family literacy and early-education context & Vocabulary learning, engagement, parent-selected targets, and design requirements for sustained home use & \cite{antony2026ella} \\
\end{xltabular}

\begin{xltabular}{\textwidth}{@{}p{0.55cm} p{3.5cm} p{4.1cm} X p{0.65cm}@{}}
\caption{Tools, rules, and divisions of labour in the primary studies.}
\label{tab:appendix-activity-organization}\\
\toprule
\textbf{ID} & \textbf{Tools} & \textbf{Rules} & \textbf{Division of labour} & \textbf{Ref.} \\
\midrule
\endfirsthead
\multicolumn{5}{@{}l}{\textit{Table~\thetable{} continued from previous page}} \\
\toprule
\textbf{ID} & \textbf{Tools} & \textbf{Rules} & \textbf{Division of labour} & \textbf{Ref.} \\
\midrule
\endhead
\midrule
\multicolumn{5}{r@{}}{\textit{Continued on next page}} \\
\endfoot
\bottomrule
\endlastfoot
P1 & LLM-personalized story-reading system & Child appropriateness, content quality, and parent guidance & System personalizes interaction; children read and respond; parents and experts evaluate & \cite{chen2025characterizing} \\
P2 & Contextual parent guidance and vocabulary-card recommendation & Balanced communication and respect for child agency & System guides parents and recommends cards; parents scaffold; children express and choose & \cite{choi2025aacesstalk} \\
P3 & Generative parent-facing spatial-language guidance & Guidance must fit play and children's needs & System suggests language; parents scaffold play; children build and communicate & \cite{liu2025bricksmart} \\
P4 & Personalized collaborative video-censorship system & Household appropriateness and negotiated media boundaries & System assesses content; parents and children establish preferences and negotiate decisions & \cite{zhao2025youthcare} \\
P5 & AI-mediated prompts embedded in media consumption & Parent oversight and developmentally appropriate reflection & System prompts reflection; children respond; parents review perceived value & \cite{shen2025easel} \\
P6 & Proposed child-centered conversational AI & Child safety, appropriate content, and parental expectations & Parents specify requirements and mediate children's prospective use & \cite{sun2024exploring} \\
P7 & LLM-supported analysis of homework interaction & NR & Parents and children conduct homework; system supports retrospective interaction analysis & \cite{gao2025homework} \\
P8 & Conversational agents & Support from knowledgeable others and child-appropriate scaffolding & Agent provides interaction; children solve problems; adults provide contextual support & \cite{figueiredo2025designing} \\
P9 & Audio capture, diarization, speech recognition, and LLM coding & Research annotation and benchmark protocols & Adults and children converse; computational tools transcribe and code; researchers evaluate & \cite{kalanadhabhatta2024playlogue} \\
P10 & LLM-generated content and educational robot & Parent review, age appropriateness, and adjustable involvement & System generates; robot interacts; parents supervise and edit; children learn & \cite{ho2025set} \\
P11 & Multi-agent story-reading system & Family guidance and appropriate critical prompts & Agents stage critique; children reason; parents participate and evaluate & \cite{wang2025charactercritique} \\
P12 & Autonomous or adult-supported storytelling robot & Experimental interaction condition and adult-support protocol & Robot leads autonomously or shares support work with an adult; children participate & \cite{olutunbi2026leads} \\
P13 & Informal AI-literacy activities and discussion materials & Shared family participation & Materials prompt inquiry; family members jointly interpret and discuss AI & \cite{long2022family} \\
P14 & Proposed AI-supported home-practice concepts & Professional boundaries and preservation of home relationships & Intended roles: AI offers guidance; parents adapt practice; professionals supply expertise & \cite{dangol2025want} \\
P15 & AI-generated artwork descriptions and explanations & Accessibility and inclusive participation & System describes; children explain; family members interpret and discuss together & \cite{chheda2025artinsight} \\
P16 & Generative writing and image tools & Adult oversight, authorship, bias, misinformation, and student agency & Systems generate and give feedback; students write; parents and teachers supervise & \cite{han2024teachers} \\
P17 & Generative image tools and multimaterial storymaking resources & Therapeutic facilitation and emotionally appropriate use & System generates material; families create; therapists facilitate interpretation & \cite{liu2024he} \\
P18 & Pervasive profiling and bespoke storybook generation & Assessment and intervention goals & System profiles and generates; children read and respond; adults support intervention & \cite{lee2024open} \\
P19 & AI-supported co-creative storytelling tools & Shared contribution and intergenerational inclusion & Grandchildren handle technology; grandparents contribute ideas; AI organizes and extends stories & \cite{kim2025bridging} \\
P20 & Co-designed parent-facing advocacy tools & IEP procedures, documentation, and children's educational rights & System supports information work; parents advocate; schools and professionals retain institutional authority & \cite{zaidi2025sociotechnical} \\
P21 & Generative AI systems used by adolescents & Age-appropriate use, safety, transparency, and parental mediation & Adolescents use systems; parents assess opportunities and risks and articulate mediation needs & \cite{eira2025parents} \\
P22 & AI-guided multimodal journaling & Privacy, adolescent agency, and appropriate caregiver access & System elicits narratives; adolescents author them; caregivers and professionals interpret & \cite{yang2026autiverse} \\
P23 & Generated questions in a co-reading interface & Questions must support meaningful, child-appropriate dialogue & System proposes questions; parents appropriate them; children respond and discuss & \cite{dietz2024contextq} \\
P24 & ChatGPT and generative visual-storytelling tools & Creativity, literacy, child safety, and adult guidance & Systems generate stories and images; adults identify requirements and mediate prospective child use & \cite{han2023design} \\
P25 & LLM-generated dialogic-reading questions & Developmental fit and parent mediation & System asks questions; children answer; parents support reading and evaluate companionship & \cite{he2025storypal} \\
P26 & LLM-augmented toys and educational drama & Family participation and appropriate anti-bullying scenarios & System and toys structure drama; children enact; parents participate and guide & \cite{liu2026dollama} \\
P27 & Multimodal LLM with observation-then-judgment prompting & Expert criteria and separation of observation from judgment & Experts define and annotate cues; system observes and judges; researchers compare alignment & \cite{shi2026towards} \\
P28 & LLM-based parental-wellbeing assistant and interaction layer & User understanding, control, reliability, and appropriate support & System responds; parents seek and revise support; co-designers shape interaction patterns & \cite{viswanathan2025interaction} \\
P29 & GenAI bedtime-story generation & Family values and emotionally appropriate representation & Remote parents provide experiences; system generates stories; children listen and interpret & \cite{xu2025accompany} \\
P30 & ARCH pediatric chatbot prototype & Clinical safety, privacy, and professional responsibility & Intended roles: chatbot supports expression and guidance; families supply context; care experts appraise the prototype & \cite{seo2025enhancing} \\
P31 & LLM-generated probes and prospective moderation controls & Household access, intervention, and data-visibility preferences & Researchers elicit parental preferences about children's chatbot use, intervention, and data visibility & \cite{driscoll2026understanding} \\
P32 & Generative AI considered in self-directed learning & Readiness, autonomy, oversight, and responsible use & Parents describe children's readiness and their mediation practices; the study examines these perspectives & \cite{xie2026understanding} \\
P33 & LLM video-appropriateness assessment & Transparency, developmental fit, and parental accountability & System assesses and explains; parents compare reasoning and make final decisions & \cite{nawshin2026well} \\
P34 & Adolescents' AI systems and literacy resources & Parental mediation practices & Adolescents develop literacy and use AI; parents mediate; study models resulting perceptions & \cite{zhang2025parental} \\
P35 & Design fictions of AI parenting assistants & Care, risk, responsibility, and acceptable intervention & Imagined systems provide support; parents interpret, accept, or resist proposed roles & \cite{petsolari2024socio} \\
P36 & AI systems considered for autistic children's support & Parental acceptance and behavioural intention & Parents report acceptance and intentions concerning children's use; these do not establish adoption & \cite{zhang2026parents} \\
P37 & Voice-enabled child-development tracking assistant & Privacy, accuracy, reminders, and caregiver coordination & System tracks and recommends; caregivers record, interpret, and coordinate care & \cite{talai2025towards} \\
P38 & LLM-adapted robot communication styles & Family health-information sensitivity and disclosure boundaries & Robot frames information; young adults disclose; parents receive and interpret health information & \cite{wester2024facing} \\
P39 & LLM-supported homeschooling platform & Family values, curriculum standards, editable drafts, and trusted-circle boundaries & System generates and compares; parents review and teach; community members share resources & \cite{rifat2026homeroom} \\
P40 & LLM-driven child-facing emotion chatbot & Child safety, privacy, empathetic guidance, and a supplementary role to parental support & System elicits and structures reflection; children identify and express emotions; parents are positioned as subsequent sources of support & \cite{seo2024chacha} \\
P41 & ChatGPT, Stable Diffusion, Midjourney, InVideo AI, and other generative tools & Child agency, shared ownership, socially appropriate participation, and avoidance of overreliance & Children co-create with peers and AI; generative tools produce multimodal content; parents evaluate risks and desired AI roles & \cite{yoo2026creativecollaboration} \\
P42 & DiSandbox, YOLOv8, EmoLLM, Midjourney, and a parent-facing app & Privacy, cultural interpretation, preliminary rather than definitive assessment, and professional oversight & Children construct sand scenes; AI guides and analyzes; parents interpret reports and communicate; experts validate and provide consultation & \cite{shi2025ineedyourhelp} \\
P43 & Culture Link image-to-image generative storytelling platform & AI-generated artifacts act as participatory prompts rather than factual records, with reciprocal family contribution & Family members contribute images and narratives; the system composes shared artifacts; children and adults jointly interpret and retell family stories & \cite{song2026ambiguousloss} \\
P44 & AI-generated social-media content and interview prompts & Youth agency, critical awareness, and responsible communication across generations & Adolescents interpret and circulate content; parents and teachers participate in intergenerational discussion & \cite{schiavo2026brainrot} \\
P45 & LLM and BERTopic case-analysis tools & Public accountability, practitioner discretion, confidentiality, and avoidance of automated overreach & Researchers analyze case records; practitioner assessments inform evaluation and interpretation & \cite{moon2026promises} \\
P46 & Envisioned generative AI household-safety agents & Household privacy, consent, authority, transparency, and differentiated responses to physical and digital threats & Family members consider prospective interventions and articulate desired boundaries and control & \cite{wen2025families} \\
P47 & ChatGPT for family-planning questions & Privacy, cultural sensitivity, comprehensibility, reproductive autonomy, and appropriate professional referral & ChatGPT supplies information; postpartum women ask and assess; family and health actors shape access and decisions & \cite{kaur2025familyplanning} \\
P48 & LLM-based conversational resilience agent & Developmentally appropriate assessment, privacy, and non-clinical support boundaries & Agent conducts reflective dialogue; children respond; adults provide the surrounding support context & \cite{hu2024grow} \\
P49 & Everyday consumer GenAI tools used in family learning & Reciprocal respect, verification, responsibility, and age-appropriate stewardship & Younger members introduce and troubleshoot tools; older members contribute judgment and context; family members learn reciprocally & \cite{wu2026warmsteward} \\
P50 & LLM-supported role-play and Socratic dialogue & Cultural fit, fairness, and child-appropriate reflection & Parents inform scenarios; system poses reflective questions; children examine and challenge stereotypes & \cite{hassan2026anyone} \\
P51 & GenAI cultural probes and prospective reflection prompter & Family values, pluralism, child-appropriate scaffolding, and avoidance of prescriptive belief change & Activities prompt discussion; children and parents reflect together; parents articulate values and design boundaries & \cite{zanardi2026dinner} \\
P52 & Generative AI systems considered for children's use & Access, perceived risk, educational value, and parental responsibility & Parents evaluate adoption and shape children's opportunities to receive AI-supported learning & \cite{baba2026digitalinequality} \\
P53 & ELLA LLM-powered social robot & Parent-selected learning targets, developmentally appropriate dialogue, bounded daily use, and home suitability & Parents select vocabulary and themes; robot generates stories and scaffolds dialogue; children interact and learn & \cite{antony2026ella} \\
\end{xltabular}


\end{document}